\documentclass[preprint,aps,prd,nofootinbib,tightenlines]{revtex4}
\usepackage[colorlinks,urlcolor=blue,linkcolor=blue,citecolor=blue]{hyperref} 
\usepackage{epsfig,graphicx,amsmath,amssymb,mathrsfs,float,color,xspace,helvet}            
\def \babar {{\it BABAR} }
\def \belle  {{Belle} }
\begin{document}
										
\title{\large{Contributions for the kaon pair from $\rho(770)$, $\omega(782)$ and their excited states in  
                  the $B\to K\bar K h$ decays} \vspace{0.2cm} }
                    
\author{Wen-Fei Wang$^{1,2}$}\email{wfwang@sxu.edu.cn}

\affiliation{$^1$Institute of Theoretical Physics, Shanxi University, Taiyuan, Shanxi 030006, China  \vspace{0.2cm}\\
                 $^2$State Key Laboratory of Quantum Optics and Quantum Optics Devices, Shanxi University, 
                        Taiyuan, Shanxi 030006, China  \vspace{0.2cm} }

\date{\today}

\begin{abstract}
We study the resonance contributions for the kaon pair originating from the intermediate states $\rho(770,1450,1700)$ 
and $\omega(782,1420,1650)$ for the three-body hadronic decays $B\to K\bar K h$ in the perturbative QCD approach, 
where $h=(\pi, K)$. The branching fractions of the virtual contributions for $K\bar K$ from the Breit-Wigner formula tails of 
$\rho(770)$ and $\omega(782)$ which have been ignored in experimental and theoretical studies for these decays are 
found larger than the corresponding contributions from the resonances $\rho(1450,1700)$ and $\omega(1420,1650)$. 
The differential branching fractions for $B\to \rho(770) h\to K\bar K h$ and $B\to\omega(782) h \to K\bar K h$ are found 
nearly unaffected by the quite different values of the full widths for $\rho(770)$ and $\omega(782)$ in this paper. The 
predictions in this work for the branching fractions of the quasi-two-body decays $B^+\to \pi^+ \rho(1450)^0\to \pi^+K^+K^-$ 
and $B^+\to \pi^+ \rho(1450)^0\to \pi^+\pi^+\pi^-$ meet the requirement of $SU(3)$ symmetry relation.
\end{abstract}   

\maketitle

\section{Introduction}\label{sec-int}  

Charmless three-body hadronic $B$ meson decays provide us a field to investigate different aspects of weak and 
strong interactions. The underlying weak decay for $b$-quark is simple which can be described well by the effective 
Hamiltonian~\cite{rmp68-1125}, but the strong dynamics in these three-body processes is very complicated, owing 
to the hadron-hadron interactions, the three-body effects~\cite{npps199-341,prd84-094001} and the rescattering 
processes~\cite{1512-09284,prd89-094013,epjc78-897,prd71-074016} in the final states, 
and also on account of the resonant contributions which are related to the scalar, vector and tensor resonances and are commonly 
described by the relativistic Breit-Wigner (BW) formula~\cite{BW-model} as well as the nonresonant contributions which are the 
rest at the amplitude level for the relevant decay processes. The experimental efforts for the three-body $B$ decays by employing 
Dalitz plot technique~\cite{prd94-1046} within the isobar formalism~\cite{pr135-B551,pr166-1731,prd11-3165} have revealed 
valuable information on involved strong and weak dynamics. But {\it a priori} model with all reliable and correct strong dynamical 
components  is needed for the Dalitz plot analyses~\cite{plb665-30}. The expressions of the decay amplitudes for those three-body decays without or have wrong factors for certain intermediate states will have negative impacts on the observables such as the 
branching fractions and $CP$ violations for the relevant decay processes.

Recently, in the amplitude analysis of the three-body decays $B^\pm\to \pi^\pm K^+K^-$, LHCb Collaboration reported 
an unexpected large fit fraction $(30.7\pm1.2\pm0.9)\%$ in Ref.~\cite{prl123-231802} for the resonance $\rho(1450)^0$ 
decaying into charged kaon pair. This fit fraction implies a branching fraction $(1.60\pm0.14)\times10^{-6}$ for the 
quasi-two-body decay $B^+\to\pi^+\rho(1450)^0 \to\pi^+ K^+K^-$~\cite{PDG-2020}, in view of the branching fractions  $(5.38\pm0.40\pm0.35)\times10^{-6}$ from Belle~\cite{prd96-031101} and $(5.0\pm0.5\pm0.5)\times10^{-6}$ presented 
by BaBar~\cite{prl99-221801} for the $B^+\to K^+K^-\pi^+$ decays. While in the $\rho$ dominant decay modes 
$B^\pm\to \pi^\pm \pi^+\pi^-$, the contribution for $\pi^+\pi^-$ pair from the intermediate state $\rho(1450)^0$ was 
found to be small but consistent with the theoretical expectation~\cite{prd96-036014} by LHCb in their recent 
works~\cite{prl124-031801,prd101-012006}. 

In Ref.~\cite{prd101-111901}, within flavour $SU(3)$ symmetry, we predicted the branching fraction for 
$B^+\to\pi^+\rho(1450)^0 \to\pi^+ K^+K^-$ to be about one tenth of that for the decay 
$B^+\to\pi^+\rho(1450)^0 \to\pi^+ \pi^+\pi^-$ and much smaller than the corresponding result in~\cite{prl123-231802,PDG-2020}, 
and our prediction got the supports from the theoretical analyses in Ref.~\cite{2007-02558}. In addition, the virtual 
contribution~\cite{plb791-342,prd69-112002,prd79-112004,prd91-092002,prd94-072001} for $K^+K^-$ from the Breit-Wigner 
(BW) formula~\cite{BW-model} tail of the resonance $\rho(770)^0$ which has been ignored by the experimental analysis was found 
to be the same order but larger than the contribution of $\rho(1450)^0\to K^+K^-$~\cite{prd101-111901}. In this work, we shall 
systematically study the contributions for the kaon pair from the resonances $\rho(770,1450,1700)$ and $\omega(782,1420,1650)$ 
in the $B\to K\bar K h$ decays within the perturbative QCD (PQCD) approach~\cite{plb504-6,prd63-054008,prd63-074009,ppnp51-85}, 
where $h$ is the bachelor state pion or kaon. As for the other $J^{PC}=1^{--}$ isovector resonances, like $\rho(1570)$, $\rho(1900)$ 
and $\rho(2150)$, we will leave their possible contributions for kaon pair to the future studies in view of their ambiguous 
nature~\cite{PDG-2020}.

The contributions for $K\bar K$ from the tails of $\rho(770)$ and $\omega(782)$ in the charmless three-body hadronic $B$ 
meson decays have been ignored in both the theoretical studies and the experimental works. But in the processes of 
$\pi^-p\to K^-K^+n$ and $\pi^+n\to K^-K^+p$~\cite{prd15-3196,prd22-2595}, $\bar p p \to K^+K^-\pi^0$~\cite{plb468-178,
epjc80-453}, $e^+e^- \to K^+K^-$~\cite{pl99b-257,pl107b-297,plb669-217,prd76-072012,prd88-032013,prd94-112006,
plb779-64,prd99-032001,zpc39-13} and $e^+ e^- \to K^0_{S}K^0_{L}$~\cite{pl99b-261,prd63-072002,plb551-27,plb760-314,
prd89-092002,jetp103-720}, the resonances $\rho(770)$ and $\omega(782)$ along with their excited states are indispensable 
for the formation of the kaon pair. In addition, the resonances $\rho(770,1450)^\pm$ are the important intermediate states for 
the $K^\pm K^0_S$ pair in the final state of hadronic $\tau$ decays~\cite{prd98-032010,prd89-072009,prd53-6037,epjc79-436}. 
The subprocesses $\rho(1450,1700)\to K\bar K$ be concerned for the decay $J/\psi\to K^+ K^-\pi^0$ in 
Refs.~\cite{prd76-094016,prd75-074017,prd95-072007,prd100-032004} could be mainly attributed to the observation of a resonant 
broad structure around $1.5$ GeV in the $K^+K^-$ mass spectrum in~\cite{prl97-142002}. While for the decays 
$B\to KKK$~\cite{prd65-092005,prd71-092003,prd74-032003,prl99-161802,prd82-073011,prd85-112010} and 
$B\to KK\pi$~\cite{prd87-091101,prl99-221801}, the unsettled $f_X(1500)$ which decaying into $K^+K^-$ channel could 
probably be related to the resonance $\rho(1450)^0$~\cite{2007-13141}.

\begin{figure}[tbp]
\centerline{\epsfxsize=14cm \epsffile{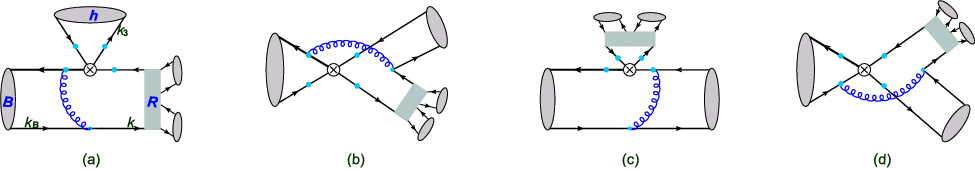}}  
\caption{Typical Feynman diagrams for the processes $B\to R h\to K\bar K h$, with $R$ represents the resonances $\rho$, 
              $\omega$ and their excited states. The dots on the quarks connecting the weak vertex $\otimes$ are the switchable 
              vertices for the hard gluons.  
              }
\label{fig-feyndiag}
\end{figure}

For the three-body decays $B\to K\bar K h$, the subprocesses $\rho\to K\bar K$ and $\omega\to K\bar K$ can not be calculated 
in the PQCD approach and will be introduced into the distribution amplitudes of the $K\bar K$ system via the kaon vector 
time-like form factors. The intermediate $\rho(770)$, $\omega(782)$ resonances and their excited states are generated in the hadronization of the light quark-antiquark pair $q\bar q^{(\prime)}$ with $q^{(\prime)}=(u, d)$ as demonstrated in the 
Fig.~\ref{fig-feyndiag} where the factorizable and nonfactorizable Feynman diagrams have been merged for the sake of simplicity. 
In the first approximation one can neglect the interaction of the $K\bar K$ pair originating from the intermediate states with the 
bachelor $h$, and study the decay processes $B\to \rho(770,1450,1700) h\to K\bar K h$ and 
$B\to\omega(782,1420,1650) h \to K\bar K h$ in the quasi-two-body framework~\cite{plb763-29,1605-03889,prd96-113003}.
The $\pi\pi\leftrightarrow KK$ rescattering effects were found have important contributions for 
$B^\pm\to \pi^\pm K^+K^-$~\cite{prl123-231802}, which would be investigated in a subsequent work. The final state 
interaction effect for the $\rho(1450,1700)\to K\bar K$ were found to be suppressed in~\cite{prd75-074017} and will be neglected 
in the numerical calculation of this work. The quasi-two-body framework based on PQCD approach has been discussed 
in detail in~\cite{plb763-29}, which has been followed in Refs.~\cite{prd101-111901,epjc80-815,jhep2003-162,prd96-036014,
prd95-056008,2007-13141,2010-12906,prd96-093011,npb923-54,epjc80-394,2102-04691} for the quasi-two-body $B$ meson 
decays in recent years.  Parallel analyses for the related three-body $B$ meson processes within QCD factorization can be found 
in Refs.~\cite{2007-08881,jhep2006-073,plb622-207,plb669-102,prd79-094005,prd72-094003,prd76-094006,prd88-114014,
prd89-074025,prd94-094015,npb899-247,2007-02558,epjc75-536,prd89-094007}, and for relevant work within the symmetries one 
is referred to Refs.~\cite{plb564-90,prd72-075013,prd72-094031,prd84-056002,plb727-136,plb726-337,prd89-074043,
plb728-579,prd91-014029}.

This paper is organized as follows. In Sec.~\ref{sec-KaonFFs}, we review the kaon vector time-like form factors, which 
are the crucial inputs for the quasi-two-body framework within PQCD and decisive for the numerical results of this work.
In Sec.~\ref{sec-kinot}, we give a brief introduction of the theoretical framework for the quasi-two-body $B$ meson 
decays within PQCD approach. In Sec.~\ref{sec-res}, we present our numerical results of the branching fractions 
and direct $CP$ asymmetries for the quasi-two-body decays $B\to \rho(770,1450,1700) h\to K\bar K h$ and 
$B\to\omega(782,1420,1650) h \to K\bar K h$, along with some necessary discussions. Summary of this work is 
given in Sec.~\ref{sec-con}. The wave functions and factorization formulae for the related decay amplitudes are 
collected in the Appendix.

\section{Kaon time-like form factors}  \label{sec-KaonFFs}   

The electromagnetic form factors for the charged and neutral kaon are important for the precise determination of the 
hadronic loop contributions to the anomalous magnetic moment of the muon and the running of the QED coupling to 
the $Z$ boson mass~\cite{prd69-093003,plb779-64,prd97-114025} and are also valuable for the measurements of the 
resonance parameters~\cite{zpc39-13,plb669-217,prd88-032013,plb779-64,prd63-072002,prd89-092002,plb760-314}. 
The kaon electromagnetic form factors have been extensively studied in Refs.~\cite{epjc79-436,prd67-034012,epjc39-41,
epjc49-697,prd81-094014} on the theoretical side. Up to now the experimental information on these form factors comes 
from the measurements of the reactions $e^+e^- \to K^+K^-$~\cite{zpc39-13,prd76-072012,prd99-032001} and 
$e^+e^- \to K^+K^-(\gamma)$~\cite{prd88-032013}.  Since $K\bar K$ is not an eigenstate of isospin, both isospin $0$ 
and $1$ resonances need to be considered in components of the form factors of kaon~~\cite{prd88-032013}. The 
combined analysis of the $e^+e^- \to K^+K^-$ and $e^+e^- \to K_SK_L$ cross sections and the spectral function in the 
$\tau^-\to K^-K^0\nu_\tau$ decay allows one to extract the isovector and isoscalar electromagnetic form factors for 
kaons~\cite{jetp129-386}.

The vector time-like form factors for charged and neutral kaons are defined by the matrix elements~\cite{zpc29-637,prd72-094003}
\begin{eqnarray} 
  \langle K^+(p_1) K^-(p_2) |\bar q\gamma_\mu (1-\gamma_5) q | 0 \rangle &=& (p_1-p_2)_\mu \,F_{K^+K^-}^{q}(s), \\
  \langle K^0(p_1)\bar K^0(p_2) |\bar q\gamma_\mu (1-\gamma_5) q | 0 \rangle &=& (p_1-p_2)_\mu\,F_{K^0\bar K^0}^{q}(s),
\end{eqnarray}
with the invariant mass square $s=p^2$ and the $K\bar K$ system momentum $p=p_1+p_2$.  These two form factors 
$F_{K^+K^-}^{q}$ and $F_{K^0\bar K^0}^{q}$ can be related to kaon electromagnetic form factors $F_{K^+}$ and $F_{K^0}$, 
which are defined by~\cite{epjc39-41}
\begin{eqnarray} 
  \langle K^+(p_1) K^-(p_2) | j^{em}_\mu | 0 \rangle &=& (p_1-p_2)_\mu \,F_{K^+}(s), \\
  \langle K^0(p_1)\bar K^0(p_2) | j^{em}_\mu | 0 \rangle &=& (p_1-p_2)_\mu\,F_{K^0}(s), 
\end{eqnarray}
and have the forms~\cite{epjc39-41}
\begin{eqnarray} 
  F_{K^+}(s)&=&+\frac12\sum_{\iota=\rho,\rho^\prime,...} c^K_\iota {\rm BW}_\iota(s) 
                    +\frac16\sum_{\varsigma=\omega,\omega^\prime,...} c^K_\varsigma {\rm BW}_\varsigma(s) 
                   +\frac13\sum_{\kappa=\phi,\phi^\prime,..} c^K_\kappa {\rm BW}_\kappa(s),  \quad
  \label{def-F-K+}   \\
  F_{K^0}(s)&=&-\frac12\sum_{\iota=\rho,\rho^\prime,...} c^K_\iota {\rm BW}_\iota(s) 
                    +\frac16\sum_{\varsigma=\omega,\omega^\prime,...} c^K_\varsigma {\rm BW}_\varsigma(s) 
                   +\frac13\sum_{\kappa=\phi,\phi^\prime,..} c^K_\kappa {\rm BW}_\kappa(s),  \quad
  \label{def-F-K0}
\end{eqnarray} 
with the electromagnetic current $j^{em}_\mu=\frac23\bar u\gamma_\mu u-\frac13\bar d\gamma_\mu d-\frac13\bar s\gamma_\mu s$ 
carried by the light quarks $u, d$ and $s$~\cite{npb250-517}. The BW formula in $F_{K^+}(s)$ and $F_{K^0}(s)$ has the 
form~\cite{zpc48-445,prd101-012006}    
\begin{eqnarray}
   {\rm BW}_R= \frac{m_{R}^2}{m_R^2-s-i m_R \Gamma_{R}(s)}\,,    
   \label{eq-BW}
\end{eqnarray}
where the $s$-dependent width is given by
\begin{eqnarray}\label{def-width}
 \Gamma_{R}(s)
             =\Gamma_R\frac{m_R}{\sqrt s} \frac{ \left| \overrightarrow{q} \right|^3}{ \left| \overrightarrow{q_0}\right|^3} 
                X^2(\left| \overrightarrow{q} \right| r^R_{\rm BW}).
  \label{eq-sdep-Gamma}
\end{eqnarray}
The Blatt-Weisskopf barrier factor~\cite{BW-X} with barrier radius $r^R_{\rm BW}=4.0$ GeV$^{-1}$~\cite{prd101-012006} 
is given by
\begin{eqnarray}
     X(z)=\sqrt{\frac{1+z^2_0}{1+z^2}}\,.
\end{eqnarray}
The magnitude of the momentum
\begin{eqnarray}
   \left| \overrightarrow{q} \right|&=&\frac{1}{2\sqrt s}\sqrt{\left[s-(m_K+m_{\bar K})^2\right]\left[s-(m_K-m_{\bar K})^2\right]}\,, 
   \label{def-q}
\end{eqnarray}
and the $\left| \overrightarrow{q_0}\right|$ is $\left| \overrightarrow{q} \right|$ at $s=m^2_R$.
One should note that $\bar c\gamma_\mu c$ can also contribute to $F_{K^+}$ and $F_{K^0}$ in the high-mass 
region~\cite{prd88-032013,prd92-054024,prd92-072008} and the BW formula for the $\rho$ family could be replaced 
with the Gounaris-Sakurai (GS) model~\cite{prl21-244} as in Refs.~\cite{epjc39-41,prd81-094014,prd86-032013}.
The $F_{K^+}$ and $F_{K^0}$ can be separated into the isospin $I=0$ and $I=1$ components as 
$F_{K^{+(0)}}=F_{K^{+(0)}}^{I=1} + F_{K^{+(0)}}^{I=0}$, with the $F_{K^+}^{I=0}=F_{K^0}^{I=0}$ and $F_{K^+}^{I=1}=-F_{K^0}^{I=1}$, 
and one has $\langle K^+(p_1) \bar{K}^0(p_2) | \bar u \gamma_\mu d | 0 \rangle =(p_1-p_2)_\mu 2F_{K^+}^{I=1}(s)$~\cite{epjc39-41,
prd96-113003}.

When concern only the contributions for ${K^+K^-}$ and ${K^0\bar K^0}$ from the resonant states $\iota=\rho(770,1450,1700)$ 
and $\varsigma=\omega(782,1420,1650)$, we have~\cite{prd72-094003} 
\begin{eqnarray} 
  F_{K^+K^-}^{u}(s)&=&F_{K^0\bar K^0}^{d}(s)= +\frac12\sum_{\iota} c^K_\iota {\rm BW}_\iota(s) 
                    +\frac12\sum_{\varsigma} c^K_\varsigma {\rm BW}_\varsigma(s),   \label{def-F-u}  \\
  F_{K^+K^-}^{d}(s)&=&F_{K^0\bar K^0}^{u}(s)=-\frac12\sum_{\iota} c^K_\iota {\rm BW}_\iota(s) 
                    +\frac12\sum_{\varsigma} c^K_\varsigma {\rm BW}_\varsigma(s).   \label{def-F-d}
\end{eqnarray}
For the $K^+\bar K^0$ and $K^0K^-$ pairs which have no contribution from the neutral resonances $\omega(782,1420,1650)$, 
we have~\cite{prd67-034012,epjc39-41,epjc79-436}
\begin{eqnarray} 
  F_{K^+\bar K^0}(s)=F_{K^0K^-}(s)=F_{K^+}(s)-F_{K^0}(s)=\sum_{\iota} c^K_\iota {\rm BW}_\iota(s).   \label{def-F-ud}
\end{eqnarray}
One should note that the different constants in Eqs.~(\ref{def-F-u})-(\ref{def-F-d}) and Eqs.~(\ref{def-F-K+})-(\ref{def-F-K0}) 
reveal the different definitions of the vector time-like and electromagnetic form factors for kaons in this work. 

\begin{table}[]   
\begin{center}
\caption{The fitted results of $c^K_R$'s in Refs.~\cite{epjc39-41,prd81-094014,jetp129-386}. 
               The column Fit-1 (Fit-2) contains the values of the constrained (unconstrained) fits. 
              }
\label{sum-cKR}   
\footnotesize 
\begin{tabular}{l l l l l l l} \hline\hline
     ~~\;$c^K_R$~          &   ~~Fit-1~\cite{epjc39-41}               & ~~Fit-2~\cite{epjc39-41}   
                                       &   ~~Fit-1~\cite{prd81-094014}        & ~~Fit-2~\cite{prd81-094014}~~    
                                       &  ~Model-I~\cite{jetp129-386}          & \,Model-II~\cite{jetp129-386}      \\
\hline  %
     \,$c^K_{\rho(770)}$   &  ~\,$1.195\pm0.009$                       &  ~\,$1.139\pm0.010$           
                                       &  ~\,$1.138\pm0.011$                       &  ~\,$1.120\pm0.007$
                                       &  ~$1.162\pm0.005$                         &  ~$1.067\pm0.041$                     \\
\,$c^K_{\omega(782)}$  &  ~\,$1.195\pm0.009$                       &  ~\,$1.467\pm0.035$           
                                       &  ~\,$1.138\pm0.011$                       &  ~\,$1.37\pm0.03$
                                       &  ~$1.26\pm0.06$                             &  ~$1.28\pm0.14$                         \\
     $c^K_{\rho(1450)}$   &  $-0.112\mp0.010$                          &  $-0.124\mp0.012$           
                                       &  $-0.043\pm0.014$                          &  $-0.107\pm0.010$
                                       &  $-0.063\pm0.014$                          &  $-0.025\pm0.008$                      \\
$c^K_{\omega(1420)}$  &  $-0.112\mp0.010$                          &  $-0.018\mp0.024$           
                                       &  $-0.043\pm0.014$                          &  $-0.173\pm0.003$
                                       &  $-0.13\pm0.03$                              &  $-0.13\pm0.02$                          \\       
     $c^K_{\rho(1700)}$   &  $-0.083\mp0.019$                          &  $-0.015\mp0.022$           
                                       &  $-0.144\pm0.015$                          &  $-0.028\pm0.012$
                                       &  $-0.160\pm0.014$                          &  $-0.234\pm0.013$                      \\
$c^K_{\omega(1650)}$  &  $-0.083\mp0.019$                          &  $-0.449\mp0.059$           
                                       &  $-0.144\pm0.015$                          &  $-0.621\pm0.020$
                                       &  $-0.37\pm0.05$                              &  $-0.234\pm0.013$                       \\                                                                                                                                                                                             
\hline\hline
\end{tabular}
\end{center}
\end{table}

The $c^K_R$ (with $R=\iota,\varsigma,\kappa$) is proportional to the coupling constant 
$g_{R K\bar K}$, and the coefficients have the constraints~\cite{jetp129-386}
\begin{eqnarray} 
\sum_{\iota=\rho,\rho^\prime,...} c^K_\iota  =1, \qquad
\frac13\sum_{\varsigma=\omega,\omega^\prime,...} c^K_\varsigma +\frac23\sum_{\kappa=\phi,\phi^\prime,..} c^K_\kappa =1
\end{eqnarray} 
to provide the proper normalizations $F_{K^+}(0)=1$ and $F_{K^0}(0)=0$, but the possibility of $SU(3)$ violations are allowed 
which will become manifest in differences between the fitted normalization coefficients~\cite{epjc39-41}.
In Refs.~\cite{epjc39-41,prd81-094014,jetp129-386}, the coefficients $c^K_R$'s for the resonances 
$\rho(770), \omega(782), \phi(1020)$ and their excited states have been fitted to the data, the results for 
$\rho(770,1450,1700)$ and $\omega(782,1420,1650)$ are summarised in Table~\ref{sum-cKR}, from which one can find that 
the fitted values for the $c^K_{\rho(1450)}$, $c^K_{\rho(1700)}$, $c^K_{\omega(1420)}$ or $c^K_{\omega(1650)}$ are quite 
different in Refs.~\cite{epjc39-41,prd81-094014,jetp129-386}. 

With the relations~\cite{epjc39-41}
\begin{eqnarray} 
    c^K_{\omega(782)}\approx\sqrt2\cdot\frac{f_{\omega(782)} g_{\omega(782)K^+ K^-}}{m_{\omega(782)}},\quad\;
    g_{\omega(782)K^+ K^-}=\frac{1}{\sqrt2}g_{\phi(1020)K^+ K^-},  \label{def-cKomega}
\end{eqnarray}
and $\Gamma_{\omega(782)\to ee}=0.60\pm0.02$ keV, $\Gamma_{\phi(1020)}=4.249\pm0.013$ MeV, the branching 
fraction $(49.2\pm0.5)\%$ for the decay $\phi(1020)\to K^+K^-$ and the masses for $K^\pm, \omega(782)$ and 
$\phi(1020)$ in~\cite{PDG-2020}, it's easy to obtain the result $1.113\pm0.019$ for the coefficient $c^K_{\omega(782)}$, 
where the error comes from the uncertainties of $\Gamma_{\omega(782)\to ee}$ and $\Gamma_{\phi(1020)}$, while the 
errors come from the uncertainties of the relevant masses are very small and have been neglected. Similarly, we have 
$c^K_{\rho(770)}=1.247\pm0.019$ with $g_{\rho(770)K^+ K^-}=g_{\omega(782)K^+ K^-}$~\cite{epjc39-41} 
and the decay constant $f_{\rho(770)}=216\pm3$ MeV~\cite{jhep1608-098}, where the error comes from the 
uncertainties of $f_{\rho(770)}$ and $\Gamma_{\phi(1020)}$. Our estimations for $c^K_{\omega(782)}$ and 
$c^K_{\rho(770)}$ are consistent with the results in~\cite{epjc39-41,prd81-094014,jetp129-386}. 
But unlike the results of Fit-2 in Refs.~\cite{epjc39-41,prd81-094014} and the values in~\cite{jetp129-386}, we have 
$c^K_{\omega(782)}$ slightly less than $c^K_{\rho(770)}$, because the decay constant (mass) for ${\omega(782)}$ is slightly 
smaller (larger) than that for ${\rho(770)}$. Supposing $f_{\rho(770)}= f_{\omega(782)}$ and $m_{\rho(770)}=m_{\omega(782)}$, 
one will have $c^K_{\omega(782)}=c^K_{\rho(770)}$ with Eq.~(\ref{def-cKomega}) and then back to the point of the 
constrained fit in~\cite{epjc39-41,prd81-094014}. To be sure, the violation of the relation 
$g_{\rho(770)K^+ K^-}=g_{\omega(782)K^+ K^-}=\frac{1}{\sqrt2}g_{\phi(1020)K^+ K^-}$ will modify our estimations for 
$c^K_{\omega(782)}$ and $c^K_{\rho(770)}$, but the violation was found quite small~\cite{plb779-64}.

In principle, the $c^K_R$ for the couplings can be calculated with the formula~\cite{prd81-094014,plb512-331}
\begin{eqnarray} 
    c^K_{R_n}=\frac{(-1)^n\Gamma(\beta^K_R-1/2)}{\alpha^\prime \sqrt\pi m^2_{R_n} \Gamma(n+1) \Gamma(\beta^K_R-1-n)}, 
     \label{fmla-cKR}
\end{eqnarray}
with $\alpha^\prime=1/(2m^2_{R_0})$, and $n=0$ for the ground states $\rho(770), \omega(782)$ and $\phi(1020)$, $n\geq1$ 
for their radial excitations. The parameters $\beta^K_R$ could be deduced from Eq.~(\ref{fmla-cKR}) with the fitted 
$c^K_{R_0}$~\cite{prd81-094014}.  With Eq.~(\ref{fmla-cKR}) one will deduce the results $c^K_{\rho(1450)}=-0.156\pm0.015$
and $c^K_{\omega(1420)}=-0.066\pm0.014$. The $c^K_{\rho(1450)}$ here is consistent with the result of Fit-2 in~\cite{epjc39-41}
but some larger than the latter for the magnitude. If we take into account the relation 
$g_{\omega(1420)K^+ K^-}\approx g_{\rho(1450)K^+ K^-}$, the big difference between $c^K_{\omega(1420)}$ and 
$c^K_{\rho(1450)}$ seems not reasonable. In view of the consistency for the coefficient $c_{\rho(1450)}$ of the pion 
electromagnetic form factor $F_\pi$ in Refs.~\cite{prd86-032013,zpc76-15,prd61-112002,pr421-191,prd78-072006} 
by different collaborations, we here propose a constraint for $c^K_{\rho(1450)}$ from the coefficient $c^\pi_{\rho(1450)}$ 
of $F_\pi$. With the relation $g_{\rho(1450)K^+ K^-}\approx \frac12 g_{\rho(1450)\pi^+ \pi^-}$ within flavour $SU(3)$ 
symmetry~\cite{epjc39-41}, one has
\begin{eqnarray} 
    |c^K_{\rho(1450)}|\approx\sqrt2\cdot\frac{|f_{\rho(1450)} g_{\rho(1450)K^+ K^-}|}{m_{\rho(1450)}}
                               \approx\frac{|f_{\rho(1450)} g_{\rho(1450)\pi^+ \pi^-}|}{\sqrt2\,m_{\rho(1450)}}
                               \approx  |c^\pi_{\rho(1450)}|,                
        \label{rela-ck1450}
\end{eqnarray}
where the different definitions for the coefficient $c^\pi_{\rho(1450)}$ in~\cite{prd86-032013,zpc76-15,prd61-112002,
pr421-191,prd78-072006} and the differences for the BW and GS models should be taken into account. In view of the results for 
$c^K_{\rho(1450)}$ in~\cite{epjc39-41} and $c^\pi_{\rho(1450)}$ in Refs.~\cite{prd86-032013,zpc76-15,prd61-112002,pr421-191,
prd78-072006}, we adopt the $c^K_{\rho(1450)}=-0.156\pm0.015$ deduced from Eq.~(\ref{fmla-cKR}) in our numerical calculation. 
In Ref.~\cite{zpc62-455}, with the analyses of the $e^+e^-$ annihilation data, $\Gamma_{\omega^\prime\to ee}$ was estimated to 
be $0.15$ keV, implies the decay constant $f_{\omega(1420)}=131$ MeV. With the $f_{\rho(1450)}=182\pm5$ MeV 
in~\cite{prd77-116009} and the masses for $\omega(1420)$ and $\rho(1450)$ in~\cite{PDG-2020}, one can estimate the ratio 
between $c^K_{\omega(1420)}$ and $c^K_{\rho(1450)}$ as $0.748\pm0.040$, then one has 
$c^K_{\omega(1420)}=-0.117\pm0.013$, which agree with the constrained result in~\cite{epjc39-41} and the corresponding values 
in~\cite{jetp129-386} as shown in Table~\ref{sum-cKR}.

The results for $c^K_{\rho(1700)}$ vary dramatically in Table~\ref{sum-cKR}, from $-0.015\mp0.022$~\cite{epjc39-41} to 
$-0.234\pm0.013$ ~\cite{jetp129-386}. A reliable reference value should come from the measurements of $F_\pi$ rather 
than the result deduced from Eq.~(\ref{fmla-cKR}) since $\rho(1700)$ is believed to be a $1^3D_1$ state in $\rho$ 
family~\cite{zpc62-455,prd55-4157,PDG-2020}. With Eq.~(\ref{rela-ck1450}) and the replacement $\rho(1450)\to\rho(1700)$ 
one has $|c^K_{\rho(1700)}|\approx0.081$ with the result $|c_{\rho^{\prime\prime}}|=0.068$ for $F_\pi$ in~\cite{prd86-032013}.
The difference between the $|c^K_{\rho(1700)}|$ and $|c_{\rho^{\prime\prime}}|$ is induced by the differences of the BW and GS 
models and the different definitions for them. Then we adopt the fitted result $-0.083\mp0.019$ for 
$c^K_{\rho(1700)}$~\cite{epjc39-41} in the numerical calculation in this work. As for the coefficient $c^K_{\omega(1650)}$, 
we employ the value $-0.083\mp0.019$ of the constrained fits in~\cite{epjc39-41} because of insufficiency of the knowledge 
for the properties of $\omega(1650)$.

\section{Kinematics and differential branching fraction}  \label{sec-kinot}   

In the light-cone coordinates, the momentum $p_B$ for the initial state $B^+, B^0$ or $B^0_s$ with the mass $m_B$ is 
written as $p_B=\frac{m_B}{\sqrt2}(1,1,0_{\rm T})$ in the rest frame of $B$ meson. In the same coordinates, the bachelor 
state pion or kaon in the concerned processes has the momentum $p_3=\frac{m_B}{\sqrt2}(1-\zeta, 0, 0_{\rm T})$, and 
its spectator quark has the momentum $k_3=(\frac{m_B}{\sqrt2}(1-\zeta)x_3, 0, k_{3{\rm T}})$. For the resonances $\rho$, 
$\omega$ and their excited states, and the $K\bar K$ system generated from them by the strong interaction, we have 
the momentum $p=\frac{m_B}{\sqrt 2}(\zeta, 1, 0_{\rm T})$ and the longitudinal polarization vector 
$\epsilon_L=\frac{1}{\sqrt 2}(-\sqrt\zeta, 1/\sqrt\zeta, 0_{\rm T})$. It's easy to check the variable $\zeta=s/m^2_B$ with the
invariant mass square $s=m^2_{K\bar K}\equiv p^2$. The spectator quark comes out from $B$ meson and goes into the 
intermediate states in hadronization shown in Fig.~\ref{fig-feyndiag}~(a) has the momenta 
$k_B=(\frac{m_B}{\sqrt2}x_B, 0, k_{B{\rm T}})$ and $k=(0, \frac{m_B}{\sqrt 2}x, k_{\rm T})$ before and after it pass through 
the hard gluon vertex. The $x_B$, $x$ and $x_3$, which run from zero to one in the numerical calculation, are the momentum 
fractions for the $B$ meson, the resonances and the bachelor final state, respectively.

For the $P$-wave $K\bar K$ system along with the subprocesses $\rho\to K\bar K$ and $\omega\to K\bar K$, the distribution 
amplitudes are organized into~\cite{prd101-111901,epjc80-815,plb763-29}
\begin{eqnarray} 
  \phi^{P\text{-wave}}_{K\bar K}(x,s)=\frac{-1}{\sqrt{2N_c}}
      \left[\sqrt{s}\,{\epsilon\hspace{-1.5truemm}/}\!_L\phi^0(x,s) 
             + {\epsilon\hspace{-1.5truemm}/}\!_L {p\hspace{-1.7truemm}/} \phi^t(x,s)   
             +\sqrt s \phi^s(x,s)  \right]\!,
\end{eqnarray}
with   
\begin{eqnarray}
   \phi^{0}(x,s)&=&\frac{3C_X F_K(s)}{\sqrt{2N_c}} x(1-x)\left[1+a_R^{0} C^{3/2}_2(1-2x) \right]\!,\label{def-DA-0}\\
   \phi^{t}(x,s)&=&\frac{3C_X F^t_K(s)}{2\sqrt{2N_c}}(1-2x)^2\left[1+a_R^t  C^{3/2}_2(1-2x)\right]\!,\label{def-DA-t}\\
   \phi^{s}(x,s)&=&\frac{3C_X F^s_K(s)}{2\sqrt{2N_c}}(1-2x)\left[1+a_R^s\left(1-10x+10x^2\right) \right]\!,\label{def-DA-s}
\end{eqnarray}
where $F_K$ is employed as the abbreviation of the vector time-like form factors in Eqs.~(\ref{def-F-u})-(\ref{def-F-ud}) and 
gain different component for different resonance contribution from to the expressions of the Eqs.~(\ref{def-F-u})-(\ref{def-F-ud}) 
in the concerned decay processes. Moreover, we have factored out the normalisation constant $C_X$ to make sure the 
the proper normalizations for the time-like form factors for kaon, and $C_X$ are given by
\begin{eqnarray}
 C_{\rho^0}=C_{\omega}=\sqrt 2, \qquad C_{\rho^\pm}=1.
\end{eqnarray} 
The Gegenbauer polynomial $C^{3/2}_2(\chi)=3\left(5\chi^2-1\right)/2$ for the distribution amplitudes $\phi^{0}$ and $\phi^{t}$, 
and the Gegenbauer moments have been catered to the data in Ref.~\cite{plb763-29} for the quasi-two-body decays 
$B\to K\rho\to K\pi\pi$. Within flavour $SU(2)$ symmetry, we adopt the same Gegenbauer moments for the $P$-wave 
$K\bar K$ system originating from the intermediate states $\omega$ and $\rho$ in this work. The vector time-like form 
factors $F^t_K$ and $F^s_K$ for the twist-$3$ distribution amplitudes are deduced from the relations 
$F^{t,s}_K(s)\approx (f^T_{\rho}/f_{\rho})F_K(s)$ and $F^{t,s}_K(s)\approx (f^T_{\omega}/f_{\omega})F_K(s)$~\cite{plb763-29} 
with the result $f^T_\rho/f_\rho=0.687$ at the scale $\mu=2$ GeV~\cite{prd78-114509}. The relation 
$f^T_\rho/f_\rho\approx f^T_\omega/f_\omega$~\cite{jhep1608-098} is employed because of the lack of a lattice QCD 
determination for $f^T_\omega$.

In PQCD approach, the factorization formula for the decay amplitude ${\mathcal A}$ of the quasi-two-body decays 
$B\to \rho h\to K\bar K h$ and $B\to \omega h\to K\bar K h$ is written as~\cite{plb561-258,prd89-074031}
\begin{eqnarray}
    {\mathcal A}=\phi_B \otimes {\mathcal H} \otimes  \phi^{P\text{-wave}}_{K\bar K} \otimes \phi_{h}   
          \label{def-DA-Q2B}
\end{eqnarray}
according to Fig.~\ref{fig-feyndiag} at leading order in the strong coupling $\alpha_s$. The hard kernel ${\mathcal H}$ here 
contains only one hard gluon exchange, and the symbol $\otimes$ means convolutions in parton momenta. For the $B$ meson 
and bachelor final state $h$ in this work, their distribution amplitudes $\phi_B$ and $\phi_{h}$ are the same as those widely 
adopted in the PQCD approach, we attach their expressions and parameters in the Appendix~\ref{sec-DAs}.  

For the $CP$ averaged differential branching fraction ($\mathcal B$), one has the 
formula~\cite{prd101-111901,prd79-094005,PDG-2020}
\begin{eqnarray}
 \frac{d{\mathcal B}}{d\zeta}=\tau_B\frac{\left| \overrightarrow{q} \right|^3 \left| \overrightarrow{q_h}\right|^3}
                                                                 {12\pi^3m^5_B}\overline{|{\mathcal A}|^2}\;,
    \label{eqn-diff-bra}
\end{eqnarray}
where $\tau_B$ is the mean lifetime for $B$ meson. The magnitude of the momentum ${\small|\overrightarrow{q_h}|}$ 
for the state $h$ in the rest frame of the intermediate states is written as
\begin{eqnarray}
   \left| \overrightarrow{q_h}\right|=\frac{1}{2\sqrt s} \sqrt{\left[m^2_{B}-(\sqrt s+m_{h})^2\right]\left[m^2_{B}-(\sqrt s-m_{h})^2\right]},  
                \label{def-qh}
\end{eqnarray}  
with $m_h$ the mass for the bachelor meson pion or kaon. When $m_K=m_{\bar K}$, the Eq.~(\ref{def-q}) has a simpler form
\begin{eqnarray}
 \left| \overrightarrow{q} \right|&=&\frac{1}{2}\sqrt{s-4m_K^2}\,.
\end{eqnarray}
Note that the cubic ${\small|\overrightarrow{q}|}$ and 
${\small|\overrightarrow{q_h}|}$ in Eq.~(\ref{eqn-diff-bra}) are caused by the introduction of the Zemach tensor 
${\small -2\overrightarrow{q}}\cdot{\small \overrightarrow{q_h}}$ which is employed to describe the angular distribution for 
the decay of spin $1$ resonances~\cite{Zemach}. The direct $CP$ asymmetry ${\mathcal A}_{CP}$ is defined as
\begin{eqnarray}
   {\mathcal A}_{CP}=\frac{{\mathcal B}(\bar B\to \bar f)-{\mathcal B}(B\to f)}{{\mathcal B}(\bar B\to \bar f)+{\mathcal B}(B\to f)}.
\end{eqnarray} 
The Lorentz invariant decay amplitudes according to Fig.~\ref{fig-feyndiag} for the decays $B\to \rho h\to K\bar K h$ and 
$B\to \omega h\to K\bar K h$ are given in the Appendix~\ref{sec-decayamp}. 

\section{Numerical results and discussions}  \label{sec-res}   

In the numerical calculation, we employ the decay constants $f_B=0.189$ GeV and $f_{B_s}=0.231$ GeV
for the $B^{0,\pm}$ and $B^0_s$ mesons~\cite{prd98-074512}, respectively, and the mean lifetimes 
$\tau_{B^0}=(1.519\pm0.004)\times 10^{-12}$~s, $\tau_{B^\pm}=(1.638\pm0.004)\times 10^{-12}$~s and 
$\tau_{B^0_s}=(1.515\pm0.004)\times 10^{-12}$~s~\cite{PDG-2020}. The masses for the relevant particles in 
the numerical calculation of this work, the full widths for the resonances $\rho(770,1450,1700)$ and 
$\omega(782,1420,1650)$, and the Wolfenstein parameters of the CKM matrix are presented in Table~\ref{params}.

\begin{table}[thb]  
\begin{center}
\caption{Masses for the relevant particles, the full widths for $\rho(770,1450,1700)$ and $\omega(782,1420,1650)$ (in units of GeV)  
               and the Wolfenstein parameters~\cite{PDG-2020}.}
\label{params}
\begin{tabular}{l}  
 \hline\hline 
  $m_{B^{0}}=5.280  \quad m_{B^{\pm}}=5.279 \qquad m_{B^0_s}\,=5.367 \quad\;\,m_{K^0}\;\,=0.498 \quad\; 
    m_{K^\pm}\;\;=0.494$ \\ 
  $m_{\pi^0}\,=0.135 \quad m_{\pi^\pm}\,=0.140 \quad m_{\rho(770)}=0.775 \quad \Gamma_{\rho(770)}=0.149 \quad\!
    m_{\omega(782)}=0.783$ \\ 
  $\Gamma_{\omega(782)}\;\,=0.00849  \qquad\qquad\;   m_{\omega(1420)}=1.410\pm0.060 \quad\;\;
      \Gamma_{\omega(1420)}=\,0.290\pm0.190$\\
  $m_{\rho(1450)}=1.465\pm0.025 \qquad\!  \Gamma_{\rho(1450)}\;=0.400\pm0.060  \quad\;\;  
    m_{\omega(1650)}\!=1.670\pm0.030$\\ 
  $\Gamma_{\omega(1650)}\,=0.315\pm0.035 \qquad\! m_{\rho(1700)}=1.720\pm0.020 \quad\;\;\, 
    \Gamma_{\rho(1700)}=0.250\pm0.100$\\
  $\lambda=0.22650\pm 0.00048  \quad\;\;  A=0.790^{+0.017}_{-0.012} \quad\;\; \bar{\rho} = 0.141^{+0.016}_{-0.017}  
      \quad\;\;\; \bar{\eta}= 0.357\pm0.01$ \\ 
 \hline\hline
\end{tabular}
\end{center}
\end{table}

\begin{table}[]   
\begin{center}
\caption{PQCD predictions of the $CP$ averaged branching fractions and the direct $CP$ asymmetries for the 
               quasi-two-body $B\to \rho(770)h \to K\bar K h$ and $B\to \omega(782)h \to K\bar K h$ decays.
               The decays with the subprocess $\rho(770)^0\to K^0\bar K^0$ or $\omega(782)\to K^0\bar K^0$ 
               have the same results as their corresponding decay modes with $\rho(770)^0\to K^+K^-$ or 
               $\omega(782)\to K^+K^-$.  }
\label{Res-770}   
\begin{tabular}{l l l} \hline\hline
\quad Decay modes       & \quad\quad${\mathcal B}$    &  \quad\quad\;${\mathcal A}_{CP}$ \\
\hline  %
 $B^+\to \pi^0 [\rho(770)^+\to] K^+\bar K^0$\;    
       &$2.01^{+0.38+0.29+0.24+0.10+0.06}_{-0.35-0.26-0.20-0.07-0.06}\times10^{-8}$ 
       &$-0.16^{+0.18+0.20+0.10+0.00}_{-0.20-0.18-0.10-0.00}$    \\
 $B^+\to \pi^+ [\rho(770)^0\to] K^+K^-$\;    
       &$1.43^{+0.26+0.19+0.11+0.06+0.04}_{-0.25-0.17-0.10-0.05-0.04}\times10^{-7}$ 
       &$-0.22^{+0.04+0.01+0.01+0.01}_{-0.04-0.01-0.01-0.01}$    \\ 
 $B^+\to\pi^+ [\omega(782)\;\to] K^+K^-$\;    
       &$4.21^{+1.67+1.03+0.08+0.21+0.14}_{-1.34-0.96-0.08-0.17-0.14}\times10^{-8}$ 
       &\;\;\,$0.02^{+0.01+0.01+0.02+0.00}_{-0.01-0.01-0.01-0.00}$    \\    
 $B^+\to K^0 [ \rho(770)^+ \to] K^+\bar K^0$\;    
       &$2.21^{+0.51+0.51+0.34+0.10+0.07}_{-0.45-0.46-0.29-0.08-0.07}\times10^{-7}$ 
       &\;\;\,$0.17^{+0.04+0.04+0.01+0.00}_{-0.05-0.03-0.02-0.00}$    \\    
 $B^+\to K^+ [\rho(770)^0 \to] K^+K^-$\;    
       &$5.15^{+0.91+0.99+0.69+0.25+0.16}_{-0.85-0.98-0.66-0.21-0.16}\times10^{-8}$ 
       &\;\;\,$0.39^{+0.03+0.04+0.04+0.00}_{-0.04-0.04-0.05-0.01}$    \\      
 $B^+\to K^+ [\omega(782)\;\to] K^+K^-$\;    
       &$8.92^{+1.67+2.33+1.19+0.43+0.30}_{-1.47-2.18-1.07-0.34-0.30}\times10^{-8}$ \;
       &\;\;\,$0.22^{+0.04+0.05+0.04+0.00}_{-0.04-0.04-0.04-0.00}$    \\                                                
\hline
 $B^0\to \pi^- [\rho(770)^+ \to] K^+\bar K^0$\;    
       &$1.02^{+0.21+0.28+0.14+0.06+0.03}_{-0.17-0.25-0.13-0.05-0.03}\times10^{-7}$ 
       &\;\;\,$0.15^{+0.04+0.04+0.00+0.00}_{-0.03-0.03-0.00-0.00}$    \\   
 $B^0\to \pi^+ [\rho(770)^-\to] K^- K^0$\;    
       &$9.59^{+3.25+1.96+0.22+0.46+0.29}_{-2.90-1.88-0.19-0.33-0.29}\times10^{-8}$ 
       &$-0.27^{+0.11+0.02+0.02+0.00}_{-0.08-0.01-0.02-0.00}$    \\                
 $B^0\to \pi^0\; [\rho(770)^0\,\to] K^+K^-$\;    
       &$1.47^{+0.96+0.53+0.19+0.13+0.04}_{-0.78-0.49-0.14-0.07-0.04}\times10^{-9}$ 
       &\;\;\,$0.19^{+0.17+0.07+0.06+0.05}_{-0.15-0.06-0.04-0.05}$    \\     
 $B^0\to \pi^0\; [\omega(782)\;\,\to] K^+K^-$\;    
       &$4.96^{+0.73+1.25+0.63+0.24+0.17}_{-0.87-1.36-0.65-0.22-0.17}\times10^{-9}$ 
       &\;\;\,$0.58^{+0.19+0.11+0.14+0.04}_{-0.18-0.11-0.14-0.04}$    \\    
 $B^0\to K^+ [\rho(770)^-\!\to] K^- K^0$\;    
       &$1.77^{+0.30+0.41+0.27+0.08+0.05}_{-0.25-0.39-0.25-0.06-0.05}\times10^{-7}$ 
       &\;\;\,$0.20^{+0.07+0.03+0.03+0.00}_{-0.08-0.02-0.03-0.00}$    \\            
 $B^0\to K^0\; [\rho(770)^0\to] K^+K^-$\;    
       &$5.44^{+0.88+1.26+0.82+0.24+0.17}_{-0.81-1.19-0.76-0.18-0.17}\times10^{-8}$ 
       &$-0.01^{+0.01+0.01+0.00+0.00}_{-0.01-0.01-0.01-0.00}$    \\      
 $B^0\to K^0\; [\omega(782)\;\to] K^+K^-$\;    
       &$5.99^{+1.15+1.60+0.88+0.22+0.20}_{-0.96-1.39-0.75-0.19-0.20}\times10^{-8}$ 
       &\;\;\,$0.01^{+0.02+0.00+0.01+0.00}_{-0.02-0.00-0.01-0.00}$    \\      
\hline  %
 $B_s^0\to\pi^- [\rho(770)^+\to] K^+\bar K^0$\;    
       &$2.31^{+0.75+0.51+0.30+0.11+0.07}_{-0.62-0.39-0.26-0.08-0.07}\times10^{-9}$ 
       &$-0.66^{+0.17+0.04+0.03+0.01}_{-0.16-0.06-0.03-0.01}$    \\    
 $B_s^0\to \pi^+ [\rho(770)^-\to] K^- K^0$\;    
       &$5.43^{+1.47+0.57+0.79+0.24+0.17}_{-1.45-0.48-0.77-0.20-0.17}\times10^{-9}$ 
       &\;\;\,$0.04^{+0.03+0.01+0.01+0.00}_{-0.04-0.01-0.01-0.00}$    \\                
 $B_s^0\to\pi^0\; [\rho(770)^0\to] K^+K^-$\;    
       &$1.63^{+0.98+0.46+0.18+0.07+0.05}_{-0.81-0.41-0.16-0.06-0.05}\times10^{-9}$ 
       &$-0.35^{+0.13+0.05+0.12+0.03}_{-0.14-0.06-0.14-0.03}$    \\    
 $B_s^0\to\pi^0\; [\omega(782)\;\to] K^+K^-$\;    
       &$8.17^{+3.83+2.37+1.22+0.51+0.28}_{-3.28-2.14-1.21-0.45-0.28}\times10^{-11}$ 
       &\;\;\,$0.11^{+0.03+0.00+0.02+0.00}_{-0.04-0.00-0.02-0.01}$    \\    
 $B_s^0\to K^- [\rho(770)^+\to] K^+ \bar K^0$\;    
       &$2.04^{+0.03+0.43+0.22+0.11+0.06}_{-0.02-0.41-0.21-0.09-0.06}\times10^{-7}$ 
       &\;\;\,$0.25^{+0.04+0.03+0.00+0.01}_{-0.04-0.03-0.00-0.01}$    \\            
 $B_s^0\to\bar K^0\; [\rho(770)^0\to] K^+K^-$\;    
       &$1.03^{+0.63+0.19+0.18+0.08+0.03}_{-0.45-0.17-0.16-0.05-0.03}\times10^{-9}$ 
       &\;\;\,$0.60^{+0.24+0.03+0.16+0.02}_{-0.22-0.04-0.14-0.04}$    \\    
 $B_s^0\to \bar K^0\; [\omega(782) \;\to] K^+K^-$\;    
       &$1.39^{+0.68+0.17+0.12+0.07+0.05}_{-0.57-0.14-0.14-0.07-0.05}\times10^{-9}$ 
       &$-0.34^{+0.29+0.06+0.01+0.03}_{-0.21-0.06-0.03-0.03}$    \\    
\hline\hline
\end{tabular}
\end{center}
\end{table}

Utilizing the differential branching fractions the Eq.~(\ref{eqn-diff-bra}) and the decay amplitudes collected in the 
Appendix~\ref{sec-decayamp}, we obtain the $CP$ averaged branching fractions and the direct $CP$ asymmetries in 
Tables~\ref{Res-770}, \ref{Res-1450}, \ref{Res-1700} for the concerned quasi-two-body decay processes 
$B\to \rho(770,1450,1700) h\to K\bar K h$ and $B\to\omega(782,1420,1650) h \to K\bar K h$. For these PQCD predictions, 
the uncertainties of the Gegenbauer moments $a_R^{0}=0.25\pm0.10$, $a_R^t=-0.50\pm0.20$ and $a_R^s=0.75\pm0.25$ 
along with the decay widths of the intermediate states contribute the first error. The second error for each result in
Tables~\ref{Res-770}, \ref{Res-1450}, \ref{Res-1700} comes from the shape parameter $\omega_B=0.40\pm0.04$ or 
$\omega_{B_s}=0.50 \pm 0.05$  in Eq.~(\ref{phib}) for the $B^{+,0}$ or $B^0_s$ meson. The third one is induced by the 
chiral scale parameters $m^h_0=\frac{m^2_{h}}{m_q+m_{q^\prime}}$ with $m^\pi_0=1.4\pm0.1$ GeV and 
$m^K_0=1.9\pm0.1$ GeV~\cite{prd76-074018} and the Gegenbauer moment $a^h_2=0.25\pm0.15$ for the bachelor 
final state pion or kaon. The fourth one comes from the Wolfenstein parameters $A$ and $\bar \rho$ listed in 
Table~\ref{params}. The uncertainties of $c^K_{\rho(770)}=1.247\pm0.019$, $c^K_{\omega(782)}=1.113\pm0.019$, 
$c^K_{\rho(1450)}=-0.156\pm0.015$, $c^K_{\omega(1420)}=-0.117\pm0.013$ and 
$c^K_{\omega(1650),\rho(1700)}=-0.083\pm0.019$ result in the fifth error for the predicted branching fractions in this work, 
while these coefficients $c^K_R$ which exist only in the kaon time-like form factors will not change the direct $CP$ asymmetries 
for the relevant decay processes. There are other errors for the PQCD predictions in this work, which come from the masses and 
the decay constants of the initial and final states, from the parameters in the distribution amplitudes for bachelor pion or kaon,
from the uncertainties of the Wolfenstein parameters $\lambda$ and $\bar{\eta}$, etc., are small and have been neglected.

\begin{table}[]   
\begin{center}
\caption{PQCD predictions of the $CP$ averaged branching ratios and the direct $CP$ asymmetries for the 
               quasi-two-body $B\to \rho(1450)h \to K\bar K h$ and $B\to \omega(1420)h \to K\bar K h$ decays.
               The decays with the subprocess $\rho(1450)^0\to K^0\bar K^0$ or $\omega(1420)\to K^0\bar K^0$ 
               have the same results as their corresponding decay modes with $\rho(1450)^0\to K^+K^-$ or 
               $\omega(1420)\to K^+K^-$.  }             
\label{Res-1450}   
\begin{tabular}{l l l} \hline\hline
\quad Decay modes       & \quad\quad${\mathcal B}$    &  \quad\quad\;${\mathcal A}_{CP}$ \\
\hline  %
 $B^+\to \pi^0 [\rho(1450)^+\to] K^+\bar K^0$\;    
       &$1.27^{+0.26+0.22+0.10+0.06+0.24}_{-0.22-0.18-0.12-0.04-0.24}\times10^{-8}$ 
       &$-0.14^{+0.24+0.21+0.11+0.00}_{-0.22-0.17-0.09-0.00}$    \\
 $B^+\to \pi^+ [\rho(1450)^0\to] K^+K^-$\;    
       &$9.46^{+1.79+1.16+0.72+0.49+1.82}_{-1.65-1.14-0.69-0.38-1.82}\times10^{-8}$ 
       &$-0.22^{+0.04+0.01+0.01+0.01}_{-0.04-0.01-0.01-0.01}$    \\
 $B^+\to\pi^+ [\omega(1420)\;\to] K^+K^-$\;    
       &$1.62^{+0.61+0.45+0.03+0.08+0.36}_{-0.52-0.39-0.02-0.07-0.36}\times10^{-8}$ 
       &\;\;\,$0.01^{+0.01+0.02+0.01+0.01}_{-0.02-0.02-0.02-0.01}$    \\ 
 $B^+\to K^0 [ \rho(1450)^+ \to] K^+\bar K^0$\;    
       &$1.20^{+0.29+0.24+0.17+0.05+0.23}_{-0.25-0.23-0.16-0.04-0.23}\times10^{-7}$ 
       &\;\;\,$0.20^{+0.04+0.03+0.02+0.00}_{-0.05-0.02-0.02-0.00}$    \\ 
 $B^+\to K^+ [\rho(1450)^0 \to] K^+K^-$\;    
       &$3.36^{+0.62+0.67+0.47+0.16+0.65}_{-0.56-0.64-0.43-0.13-0.65}\times10^{-8}$ 
       &\;\;\,$0.42^{+0.03+0.04+0.05+0.01}_{-0.03-0.03-0.05-0.01}$    \\    
 $B^+\to K^+ [\omega(1420)\;\to] K^+K^-$\;    
       &$3.09^{+0.64+0.80+0.42+0.15+0.69}_{-0.57-0.73-0.37-0.12-0.69}\times10^{-8}$ 
       &\;\;\,$0.32^{+0.05+0.05+0.03+0.01}_{-0.05-0.05-0.03-0.01}$    \\                                            
\hline
 $B^0\to \pi^- [\rho(1450)^+ \to] K^+\bar K^0$\;    
       &$7.39^{+1.58+2.20+1.01+0.41+1.42}_{-1.31-1.86-0.96-0.33-1.42}\times10^{-8}$ 
       &\;\;\,$0.16^{+0.02+0.05+0.01+0.01}_{-0.03-0.03-0.01-0.01}$    \\        
 $B^0\to \pi^+ [\rho(1450)^-\to] K^- K^0$\;    
       &$6.94^{+2.04+1.40+0.14+0.33+1.33}_{-1.94-1.38-0.14-0.25-1.33}\times10^{-8}$ 
       &$-0.27^{+0.12+0.02+0.02+0.00}_{-0.08-0.01-0.02-0.00}$    \\                       
 $B^0\to \pi^0\; [\rho(1450)^0\,\to] K^+K^-$\;    
       &$8.48^{+5.96+3.07+0.81+0.68+1.63}_{-5.14-3.01-0.78-0.49-1.63}\times10^{-10}$ 
       &\;\;\,$0.20^{+0.21+0.10+0.09+0.06}_{-0.17-0.08-0.07-0.05}$    \\        
 $B^0\to \pi^0\; [\omega(1420)\;\,\to] K^+K^-$\;    
       &$2.08^{+0.32+0.58+0.28+0.10+0.46}_{-0.37-0.66-0.32-0.08-0.46}\times10^{-9}$ 
       &\;\;\,$0.58^{+0.17+0.10+0.11+0.02}_{-0.16-0.11-0.09-0.02}$    \\        
 $B^0\to K^+ [\rho(1450)^-\!\to] K^- K^0$\;    
       &$1.18^{+0.20+0.27+0.18+0.05+0.23}_{-0.17-0.25-0.17-0.04-0.23}\times10^{-7}$ 
       &\;\;\,$0.22^{+0.08+0.03+0.04+0.00}_{-0.08-0.02-0.04-0.00}$    \\                
 $B^0\to K^0\; [\rho(1450)^0\to] K^+K^-$\;    
       &$3.69^{+0.67+0.84+0.55+0.16+0.71}_{-0.60-0.82-0.51-0.12-0.71}\times10^{-8}$ 
       &$-0.01^{+0.01+0.01+0.00+0.00}_{-0.02-0.01-0.01-0.00}$    \\            
$B^0\to K^0\; [\omega(1420)\;\to] K^+K^-$\;    
       &$2.07^{+0.48+0.48+0.29+0.08+0.46}_{-0.46-0.45-0.26-0.06-0.46}\times10^{-8}$ 
       &$-0.02^{+0.04+0.03+0.01+0.00}_{-0.02-0.03-0.01-0.00}$    \\         
\hline  %
 $B_s^0\to\pi^- [\rho(1450)^+\to] K^+\bar K^0$\;    
       &$1.55^{+0.39+0.30+0.16+0.07+0.30}_{-0.33-0.28-0.14-0.05-0.30}\times10^{-9}$ 
       &$-0.66^{+0.15+0.04+0.05+0.02}_{-0.16-0.08-0.04-0.01}$    \\        
 $B_s^0\to \pi^+ [\rho(1450)^-\to] K^- K^0$\;    
       &$4.54^{+1.30+0.37+0.69+0.20+0.87}_{-1.27-0.40-0.67-0.16-0.87}\times10^{-9}$ 
       &\;\;\,$0.04^{+0.03+0.01+0.02+0.00}_{-0.05-0.01-0.01-0.00}$    \\                
 $B_s^0\to\pi^0\; [\rho(1450)^0\to] K^+K^-$\;    
       &$1.15^{+0.72+0.35+0.09+0.05+0.22}_{-0.59-0.30-0.12-0.04-0.22}\times10^{-9}$ 
       &$-0.36^{+0.12+0.05+0.10+0.02}_{-0.16-0.04-0.14-0.03}$    \\        
 $B_s^0\to\pi^0\; [\omega(1420)\;\to] K^+K^-$\;    
       &$3.67^{+1.59+1.17+0.65+0.21+0.82}_{-1.38-0.97-0.58-0.19-0.82}\times10^{-11}$ 
       &\;\;\,$0.14^{+0.03+0.00+0.01+0.00}_{-0.02-0.01-0.01-0.01}$    \\           
 $B_s^0\to K^- [\rho(1450)^+\to] K^+ \bar K^0$\;    
       &$1.49^{+0.07+0.31+0.16+0.08+0.29}_{-0.06-0.30-0.15-0.06-0.29}\times10^{-7}$ 
       &\;\;\,$0.25^{+0.04+0.03+0.00+0.01}_{-0.04-0.03-0.00-0.01}$    \\                
 $B_s^0\to\bar K^0\; [\rho(1450)^0\to] K^+K^-$\;    
       &$6.86^{+4.09+0.81+1.03+0.44+1.32}_{-3.56-0.75-0.94-0.39-1.32}\times10^{-10}$ 
       &\;\;\,$0.64^{+0.29+0.02+0.08+0.05}_{-0.27-0.01-0.12-0.07}$    \\        
 $B_s^0\to \bar K^0\; [\omega(1420) \;\to] K^+K^-$\;    
       &$5.79^{+3.28+0.53+0.63+0.34+1.29}_{-2.39-0.42-0.57-0.31-1.29}\times10^{-10}$ 
       &$-0.54^{+0.29+0.13+0.05+0.01}_{-0.33-0.12-0.05-0.03}$    \\        
\hline\hline
\end{tabular}
\end{center}
\end{table}

\begin{table}[]   
\begin{center}
\caption{PQCD predictions of the $CP$ averaged branching ratios and the direct $CP$ asymmetries for the 
               quasi-two-body $B\to \rho(1700)h \to K\bar K h$ and $B\to \omega(1650)h \to K\bar K h$ decays.
               The decays with the subprocess $\rho(1700)^0\to K^0\bar K^0$ or $\omega(1650)\to K^0\bar K^0$ 
               have the same results as their corresponding decay modes with $\rho(1700)^0\to K^+K^-$ or 
               $\omega(1650)\to K^+K^-$.  }                  
\label{Res-1700}   
\begin{tabular}{l l l} \hline\hline
\quad Decay modes       & \quad\quad${\mathcal B}$    &  \quad\quad\;${\mathcal A}_{CP}$ \\
\hline  %
 $B^+\to \pi^0 [\rho(1700)^+\to] K^+\bar K^0$\;    
       &$1.03^{+0.21+0.20+0.09+0.05+0.47}_{-0.18-0.17-0.10-0.04-0.47}\times10^{-8}$ 
       &$-0.15^{+0.22+0.23+0.13+0.01}_{-0.23-0.21-0.12-0.00}$    \\
 $B^+\to \pi^+ [\rho(1700)^0\to] K^+K^-$\;    
       &$8.71^{+1.47+1.20+0.61+0.46+3.99}_{-1.34-1.17-0.59-0.36-3.99}\times10^{-8}$ 
       &$-0.25^{+0.03+0.02+0.01+0.01}_{-0.03-0.01-0.01-0.01}$    \\
 $B^+\to\pi^+ [\omega(1650)\;\to] K^+K^-$\;    
       &$1.48^{+0.42+0.32+0.02+0.01+0.68}_{-0.34-0.28-0.02-0.01-0.68}\times10^{-9}$ 
       &\;\;\,$0.02^{+0.01+0.00+0.00+0.00}_{-0.01-0.00-0.00-0.00}$    \\ 
 $B^+\to K^0 [ \rho(1700)^+ \to] K^+\bar K^0$\;    
       &$1.08^{+0.27+0.21+0.18+0.05+0.49}_{-0.25-0.19-0.15-0.03-0.49}\times10^{-7}$ 
       &\;\;\,$0.21^{+0.05+0.04+0.03+0.00}_{-0.06-0.03-0.02-0.00}$    \\ 
 $B^+\to K^+ [\rho(1700)^0 \to] K^+K^-$\;    
       &$2.85^{+0.50+0.49+0.35+0.14+1.30}_{-0.49-0.48-0.32-0.11-1.30}\times10^{-8}$ 
       &\;\;\,$0.47^{+0.02+0.04+0.05+0.01}_{-0.02-0.03-0.05-0.01}$    \\    
 $B^+\to K^+ [\omega(1650)\;\to] K^+K^-$\;    
       &$2.81^{+0.53+0.66+0.36+0.13+1.29}_{-0.47-0.59-0.32-0.10-1.29}\times10^{-8}$ 
       &\;\;\,$0.36^{+0.03+0.05+0.05+0.01}_{-0.04-0.05-0.05-0.01}$    \\                                            
\hline
 $B^0\to \pi^- [\rho(1700)^+ \to] K^+\bar K^0$\;    
       &$4.38^{+0.80+1.17+0.50+0.23+2.01}_{-0.73-1.06-0.48-0.19-2.01}\times10^{-8}$ 
       &\;\;\,$0.18^{+0.03+0.03+0.01+0.01}_{-0.02-0.03-0.01-0.01}$    \\        
 $B^0\to \pi^+ [\rho(1700)^-\to] K^- K^0$\;    
       &$6.66^{+1.78+1.41+0.13+0.32+3.05}_{-1.69-1.40-0.12-0.24-3.05}\times10^{-8}$ 
       &$-0.29^{+0.12+0.02+0.02+0.01}_{-0.08-0.02-0.02-0.01}$    \\                       
 $B^0\to \pi^0\; [\rho(1700)^0\,\to] K^+K^-$\;    
       &$8.11^{+5.46+3.02+0.82+0.68+3.71}_{-4.98-2.97-0.80-0.54-3.71}\times10^{-10}$ 
       &\;\;\,$0.18^{+0.20+0.08+0.07+0.04}_{-0.18-0.06-0.07-0.04}$    \\        
 $B^0\to \pi^0\; [\omega(1650)\;\,\to] K^+K^-$\;    
       &$1.48^{+0.31+0.44+0.15+0.06+0.68}_{-0.34-0.39-0.16-0.06-0.68}\times10^{-9}$ 
       &\;\;\,$0.57^{+0.21+0.07+0.09+0.01}_{-0.17-0.09-0.07-0.01}$    \\        
 $B^0\to K^+ [\rho(1700)^-\!\to] K^- K^0$\;    
       &$9.95^{+1.87+1.83+1.31+0.44+4.56}_{-1.60-1.61-1.15-0.32-4.56}\times10^{-8}$ 
       &\;\;\,$0.28^{+0.07+0.01+0.05+0.00}_{-0.09-0.01-0.04-0.00}$    \\                
 $B^0\to K^0\; [\rho(1700)^0\to] K^+K^-$\;    
       &$2.94^{+0.54+0.57+0.38+0.13+1.35}_{-0.53-0.56-0.36-0.09-1.35}\times10^{-8}$ 
       &$-0.01^{+0.01+0.00+0.01+0.00}_{-0.01-0.00-0.01-0.00}$    \\            
 $B^0\to K^0\; [\omega(1650)\;\to] K^+K^-$\;    
       &$1.89^{+0.43+0.39+0.22+0.08+0.87}_{-0.38-0.36-0.19-0.07-0.87}\times10^{-8}$ 
       &$-0.01^{+0.04+0.00+0.01+0.00}_{-0.03-0.00-0.00-0.00}$    \\         
\hline  %
 $B_s^0\to\pi^- [\rho(1700)^+\to] K^+\bar K^0$\;    
       &$1.37^{+0.34+0.29+0.14+0.06+0.63}_{-0.31-0.27-0.14-0.05-0.63}\times10^{-9}$ 
       &$-0.70^{+0.16+0.04+0.01+0.01}_{-0.15-0.07-0.04-0.01}$    \\        
 $B_s^0\to \pi^+ [\rho(1700)^-\to] K^- K^0$\;    
       &$3.57^{+0.94+0.30+0.54+0.16+1.63}_{-0.86-0.32-0.52-0.13-1.63}\times10^{-9}$ 
       &\;\;\,$0.07^{+0.04+0.01+0.02+0.00}_{-0.05-0.02-0.02-0.00}$    \\                
 $B_s^0\to\pi^0\; [\rho(1700)^0\to] K^+K^-$\;    
       &$1.01^{+0.59+0.35+0.09+0.04+0.46}_{-0.51-0.30-0.11-0.03-0.46}\times10^{-9}$ 
       &$-0.29^{+0.11+0.06+0.12+0.01}_{-0.18-0.08-0.15-0.01}$    \\        
 $B_s^0\to\pi^0\; [\omega(1650)\;\to] K^+K^-$\;    
       &$3.14^{+1.35+1.10+0.53+0.19+1.44}_{-1.29-0.98-0.49-0.16-1.44}\times10^{-11}$ 
       &\;\;\,$0.15^{+0.06+0.02+0.02+0.01}_{-0.05-0.01-0.03-0.01}$    \\           
 $B_s^0\to K^- [\rho(1700)^+\to] K^+ \bar K^0$\;    
       &$1.14^{+0.07+0.25+0.12+0.06+0.52}_{-0.07-0.24-0.12-0.05-0.52}\times10^{-7}$ 
       &\;\;\,$0.29^{+0.04+0.04+0.01+0.01}_{-0.04-0.03-0.01-0.01}$    \\                
 $B_s^0\to\bar K^0\; [\rho(1700)^0\to] K^+K^-$\;    
       &$4.21^{+1.90+0.47+0.55+0.29+1.93}_{-1.70-0.42-0.50-0.26-1.93}\times10^{-10}$ 
       &\;\;\,$0.67^{+0.25+0.03+0.12+0.04}_{-0.26-0.02-0.16-0.03}$    \\        
 $B_s^0\to \bar K^0\; [\omega(1650) \;\to] K^+K^-$\;    
       &$4.18^{+1.44+0.42+0.50+0.27+1.91}_{-1.17-0.38-0.43-0.23-1.91}\times10^{-10}$ 
       &$-0.64^{+0.26+0.08+0.09+0.03}_{-0.19-0.08-0.12-0.05}$    \\        
\hline\hline
\end{tabular}
\end{center}
\end{table}

The PQCD predictions are omitted in Tables~\ref{Res-770}, \ref{Res-1450}, \ref{Res-1700} for those quasi-two-body decays with 
the subprocesses $\rho(770, 1450, 1700)^0 \to K^0\bar K^0$ and $\omega(782, 1420, 1650)\to K^0\bar K^0$ . The variations 
caused by the small mass difference between $K^\pm$ and $K^0$ for the branching fraction and direct $CP$ asymmetry of a 
decay mode with one of these intermediate states decaying into $K^0\bar K^0$ or $K^+K^-$ are tiny. 
As the examples, we calculate the the branching fractions for the decays $B^+\to \pi^+ \rho(770)^0$, $B^+\to K^+ \rho(770)^0$, 
$B^+\to \pi^+ \omega(782)$ and $B^+\to K^+ \omega(782)$ with the resonances $\rho(770)^0$ and $\omega(782)$ decay into 
the final state $K^0\bar K^0$. Their four branching fractions with the same sources for the errors as these results in 
Table~\ref{Res-770} are predicted to be 
\begin{eqnarray}
{\mathcal B}(B^+\to \pi^+ \rho(770)^0\to \pi^+ K^0\bar K^0)
                    &=& 1.40^{+0.26+0.17+0.10+0.06+0.04}_{-0.24-0.17-0.10-0.06-0.04} \times 10^{-7},   \\ 
{\mathcal B}(B^+\to K^+ \rho(770)^0 \to K^+  K^0\bar K^0)
                    &=& 5.08^{+0.92+1.00+0.70+0.25+0.15}_{-0.83-0.97-0.65-0.20-0.15} \times 10^{-8},     \\ 
{\mathcal B}(B^+\to \pi^+ \omega(782)\to \pi^+ K^0\bar K^0)
                    &=& 4.14^{+1.64+1.02+0.07+0.20+0.14}_{-1.32-0.94-0.08-0.16-0.14} \times 10^{-8},    \\ 
{\mathcal B}(B^+\to K^+ \omega(782) \to K^+ K^0\bar K^0)
                    &=& 8.79^{+1.65+2.30+1.17+0.42+0.30}_{-1.44-2.15-1.03-0.33-0.30} \times 10^{-8}.         
\end{eqnarray}
It's easy to check that these branching fractions are very close to the results in Table~\ref{Res-770} for the corresponding 
decay modes with $\rho(770)^0$ and $\omega(782)$ decaying into $K^+K^-$. The impacts from the mass difference 
of $K^\pm$ and $K^0$ for the direct $CP$ asymmetries for the relevant processes are even smaller, which could be inferred 
from the comparison of the results in Table~\ref{Res-770} with 
\begin{eqnarray}
{\mathcal A}_{CP}(B^+\to \pi^+ \rho(770)^0\to \pi^+ K^0\bar K^0)
                   &=& -0.22^{+0.04+0.01+0.01+0.01}_{-0.04-0.01-0.01-0.01},    \\    
{\mathcal A}_{CP}(B^+\to K^+ \rho(770)^0 \to K^+  K^0\bar K^0)
                   &=& \;\;\;0.39^{+0.03+0.04+0.04+0.01}_{-0.04-0.04-0.04-0.01}.         
\end{eqnarray}
For the decay modes $B^+\to \pi^+ \rho(1450)^0$ and $B^+\to K^+ \rho(1450)^0$ with $\rho(1450)^0\to K^0\bar K^0$, 
we have the central values $9.32\times10^{-8}$ and $-0.22$, $3.30\times10^{-8}$ and $0.42$ as their branching fractions and 
direct $CP$ asymmetries, respectively, which are also very close to the results in Table~\ref{Res-1450} for the corresponding decay 
processes with $\rho(1450)^0\to K^+K^-$. In view of the large errors for the predictions in Tables~\ref{Res-770}, \ref{Res-1450}, 
\ref{Res-1700}, we set the concerned decays with the subprocess $\rho(770, 1450, 1700)^0\to K^0\bar K^0$ or 
$\omega(782, 1420, 1650)\to K^0\bar K^0$ have the same results as their corresponding decay modes with the resonances 
decaying into $K^+K^-$. It should be stressed that the $K^0\bar K^0$ with the $P$-wave resonant origin in the final state of 
$B\to K\bar K h$ decays can not generate the $K^0_SK^0_S$ system because of the Bose-Einstein statistics.

From the branching fractions in Tables~\ref{Res-770}, \ref{Res-1450}, one can find that the virtual contributions for $K\bar K$ 
from the BW tails of the intermediate states $\rho(770)$ and $\omega(782)$ in those quasi-two-body decays which have 
been ignored in experimental and theoretical studies are all larger than the corresponding results from $\rho(1450)$ and 
$\omega(1420)$. Specifically, the branching fractions in Table~\ref{Res-770} with the resonances $\rho(770)^0$ and 
$\rho(770)^\pm$ are about $1.2$-$1.8$ times of the corresponding results in Table~\ref{Res-1450} for the decays with 
$\rho(1450)^0$ and $\rho(1450)^\pm$, while the six predictions for the branching fractions in Table~\ref{Res-770} with 
$\omega(782)$ in the quasi-two-body decay processes are about $2.2$-$2.9$ times of the corresponding values for the 
decays with the resonance $\omega(1420)$ in Table~\ref{Res-1450}. The difference of the multiples between the results 
of the branching fractions with the resonances $\rho$ and $\omega$ in Table~\ref{Res-770} and Table~\ref{Res-1450} 
should mainly be attributed to the relatively small value for the $c^K_{\omega(1420)}$ adopted in this work comparing with 
$c^K_{\rho(1450)}$. 

\begin{figure}[tbp]
\centerline{\epsfxsize=9cm \epsffile{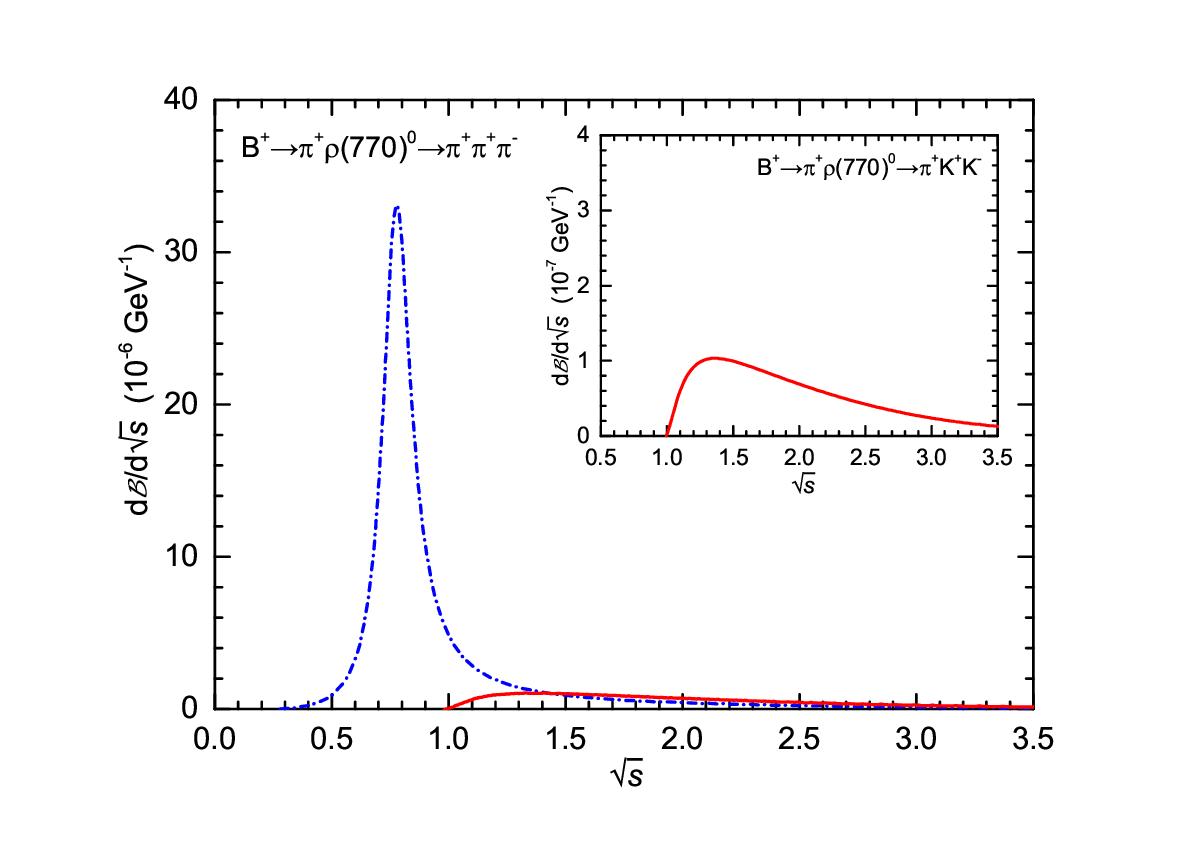}   \hspace{-0.5cm}
                  \epsfxsize=9cm \epsffile{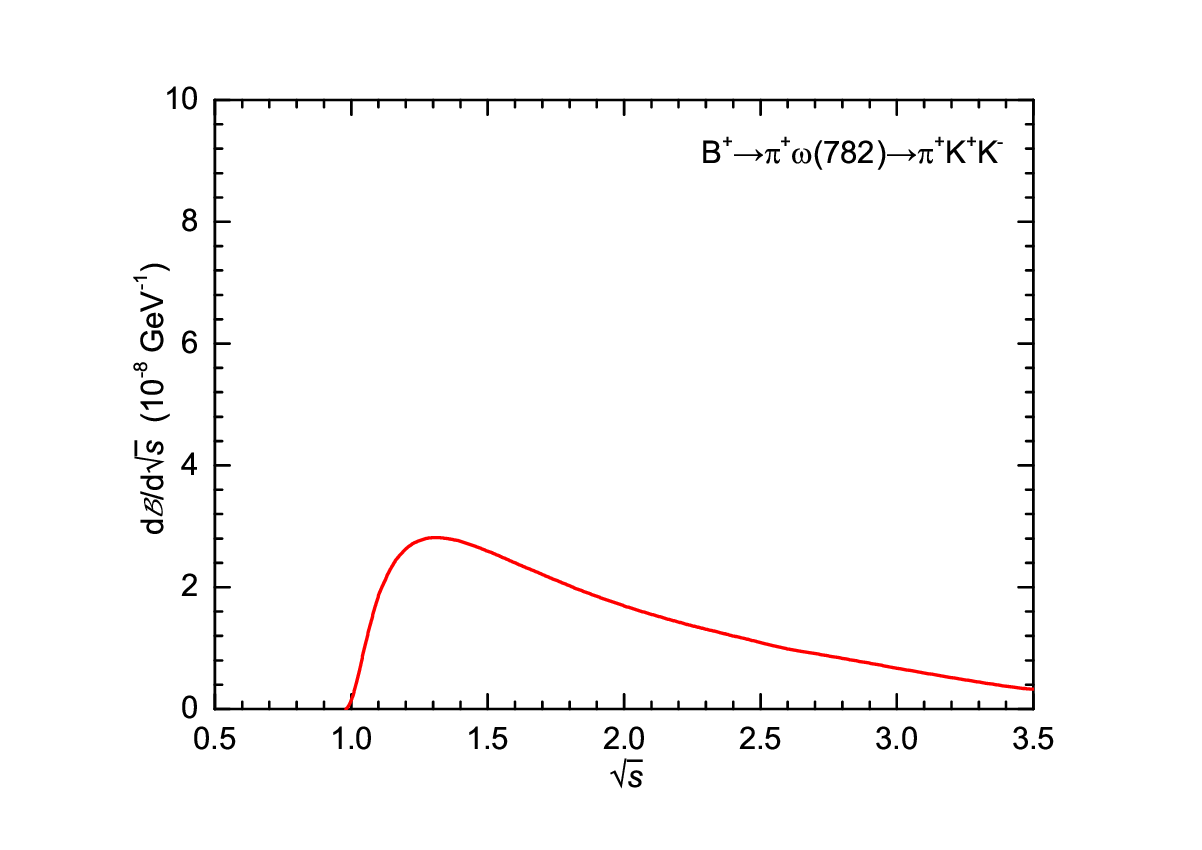}}  
\caption{The differential branching fractions for the decays $B^+\to \pi^+ [\rho(770)^0\to] K^+K^-$ (left) and 
               $B^+\to \pi^+ [\omega(782)\to] K^+K^-$ (right). The big diagram in the left is for the comparison 
               for the differential branching fractions of $B^+\to \pi^+ [\rho(770)^0\to] K^+K^-$ and 
               $B^+\to \pi^+ [\rho(770)^0\to] \pi^+\pi^-$, in which the solid line for $B^+\to \pi^+ [\rho(770)^0\to] K^+K^-$ 
               is magnified by a factor of $10$.
              }
\label{fig-dB-770}
\end{figure}

It is remarkable for these virtual contributions in Table~\ref{Res-770} that their differential branching fractions are nearly 
unaffected by the full widths of $\rho(770)$ and $\omega(782)$, which could be concluded from the Fig.~\ref{fig-dB-770}.
In this figure, the lines in the left diagram for $B^+\to \pi^+ [\rho(770)^0\to] K^+K^-$ and in the right diagram for 
$B^+\to \pi^+ [\omega(782)\to] K^+K^-$ have very similar shape although there is a big difference between the values for 
the widths of $\rho(770)$ and $\omega(782)$ as listed in Table~\ref{params}. The best explanation for Fig.~\ref{fig-dB-770} is 
that the imaginary part of the denominator in the BW formula the Eq.~(\ref{eq-BW}) which hold the energy dependent width 
for the resonances $\rho(770)$ or $\omega(782)$ becomes unimportant when the invariant mass square $s$ is large enough
even if one employs the effective mass defined by the {\it ad hoc} formula~\cite{prd91-092002,prd90-072003} to replace 
the $m^2_R$ in $\left| \overrightarrow{q_0}\right|$ in Eq.~(\ref{eq-sdep-Gamma}) or calculates the energy dependent width 
with the partial widths and the branching ratios for the intermediate state as in Refs.~\cite{plb779-64,prd88-032013,
plb760-314,prd76-072012}. At this point, the BW expression for $\rho(770)$ or $\omega(782)$ is charged by 
the coefficient $c^K_R$ in the time-like form factors for kaons and the gap between the invariant mass square $s$ 
for kaon pair and the squared mass of the resonance. Although the threshold of kaon pair is not far from the pole masses 
of $\rho(770)$ and $\omega(782)$, thanks to the strong suppression from the factor $\left| \overrightarrow{q}\right|^3$ in 
Eq.~(\ref{eqn-diff-bra}), the differential branching fractions for those processes with $\rho(770)$ or $\omega(782)$ decaying 
into kaon pair will reach their peak at about $1.35$ GeV as shown in Fig.~\ref{fig-dB-770}.

\begin{figure}[tbp]
\centerline{\epsfxsize=10.0cm \epsffile{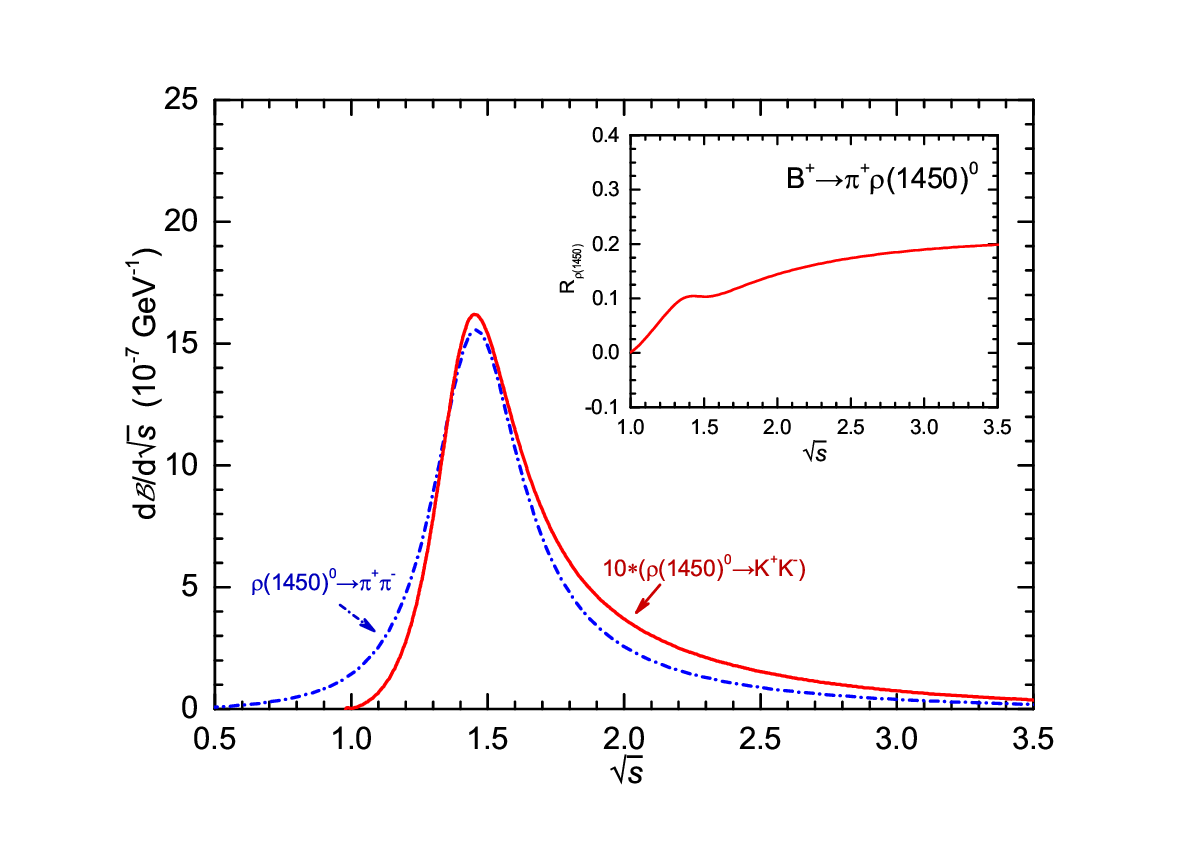}}  
\caption{The differential branching fractions for the decays $B^+\to \pi^+ [\rho(1450)^0\to] K^+K^-$ (solid line) which 
                is magnified by a factor of $10$ and $B^+\to \pi^+ [\rho(1450)^0\to] \pi^+\pi^-$(dash-dot line) in the large diagram
                and curve for the $\sqrt s$ dependent ratio $R_{\rho(1450)}$ in the small one.
              }
\label{fig-dep1450}
\end{figure}

As we have stated in Ref.~\cite{prd101-111901}, the bumps in Fig.~\ref{fig-dB-770} for $B^+\to \pi^+ [\rho(770)^0\to] K^+K^-$ 
and $B^+\to \pi^+ [\omega(782)^0\to] K^+K^-$ are generated by the tails of the BW formula for the resonances $\rho(770)$ and 
$\omega(782)$ along with the phase space factors in Eq.~(\ref{eqn-diff-bra}) and should not be taken as the evidence for a new 
resonant state at about $1.35$ GeV. When we compare the curves for the differential branching fractions for 
$B^+\to \pi^+ [\rho(770)^0\to] K^+K^-$ and $B^+\to \pi^+ [\rho(770)^0\to] \pi^+\pi^-$, we can understand this point well.  
In order to make a better contrast, the differential branching fraction for $B^+\to \pi^+ [\rho(770)^0\to] K^+K^-$ is magnified $10$ 
times in the big one of the left diagram of Fig.~\ref{fig-dB-770}. The dash-dot line for $B^+\to \pi^+ [\rho(770)^0\to] \pi^+\pi^-$ 
shall climb to its peak at about the pole mass of $\rho(770)^0$ and then descend as exhibited in Fig.~\ref{fig-dB-770}. While this 
pattern is inapplicable for the decay process of $B^+\to \pi^+ [\rho(770)^0\to] K^+K^-$, its curve can only show the existence 
from the threshold of kaon pair where the $\sqrt s$ has already crossed the peak of BW for $\rho(770)^0$. As $\sqrt s$ becoming 
larger, the effect of the full width for $\rho(770)$ fade from the stage, the ratio between the differential branching fractions for the 
quasi-two-body decays $B^+\to \pi^+ [\rho(770)^0\to] K^+K^-$ and $B^+\to \pi^+ [\rho(770)^0\to] \pi^+\pi^-$ will tend to 
be a constant which is proportional to the value of $|g_{\rho(770)K^+ K^-}/g_{\rho(770)\pi^+\pi^-}|^2$ if the phase space for 
the decay process is large enough. This conclusion can also be demonstrated well from the curve of the ratio 
\begin{eqnarray}
   R_{\rho(1450)}(\sqrt s)=\frac{d{\mathcal B}(B^+\to \pi^+ [\rho(1450)^0\to] K^+K^-)/d\sqrt{s}}
                                   {d{\mathcal B}(B^+\to \pi^+ [\rho(1450)^0\to] \pi^+\pi^-)/d\sqrt{s}}
   \label{def-R-1450}
\end{eqnarray}
for the decays $B^+\to \pi^+ [\rho(1450)^0\to] K^+K^-$ and $B^+\to \pi^+ [\rho(1450)^0\to] \pi^+\pi^-$ in Fig.~\ref{fig-dep1450}.
The solid line which stands for the $B^+\to \pi^+ [\rho(1450)^0\to] K^+K^-$ decay and has been magnified $10$ times will arise at 
the threshold of kaon pair in Fig.~\ref{fig-dep1450} and contribute the zero for $R_{\rho(1450)}$ because of the factor 
$\left| \overrightarrow{q}\right|^3$ in Eq.~(\ref{eqn-diff-bra}), and the following for $R_{\rho(1450)}$ is a rapid rise to the value 
about $0.1$ in the region where the main portion of the branching fractions for $B^+\to \pi^+ [\rho(1450)^0\to] K^+K^-$ and 
$B^+\to \pi^+ [\rho(1450)^0\to] \pi^+\pi^-$ concentrated, then $R_{\rho(1450)}$ is going to the value 
$|g_{\rho(1450)K^+ K^-}/g_{\rho(1450)\pi^+\pi^-}|^2$ as the rise of $s$.

With the help of the factorization relation 
$\Gamma(B^+\to \rho(1450)^0\pi^+\to h^+h^-\pi^+)\approx\Gamma(B^+\to \rho(1450)^0\pi^+){\mathcal B}(\rho(1450)^0\to h^+h^-)$ 
\cite{2011-03201,2011-07468}, the ratio $R_{\rho(1450)}$ can be related to the the coupling constants $g_{\rho(1450)^0\pi^+\pi^-}$ 
and $g_{\rho(1450)^0K^+K^-}$ with the expression
\begin{eqnarray}
  g_{\rho(1450)^0h^+h^-}=\sqrt{\frac{6\pi m^2_{\rho(1450)}\Gamma_{\rho(1450)}{\mathcal B}_{\rho(1450)^0\to h^+h^-}}{q^3}}\,,
  \label{def-cocont}
\end{eqnarray}
here $q=\frac{1}{2}\sqrt{m^2_{\rho(1450)}-4m_h^2}$ and $h$ is pion or kaon. Utilizing the relation 
$g_{\rho(1450)^0K^+K^-}\approx \frac12 g_{\rho(1450)^0\pi^+\pi^-}$~\cite{epjc39-41} one has~\cite{prd101-111901}
\begin{eqnarray}
  R_{\rho(1450)} &=& \frac{{\mathcal B}(\rho(1450)^0\to K^+ K^-)}{{\mathcal B}(\rho(1450)^0 \to\pi^+\pi^-)} 
                           \approx \frac{g^2_{\rho(1450)^0K^+K^-} (m^2_{\rho(1450)}-4m^2_K)^{3/2}}
                                                {g^2_{\rho(1450)^0\pi^+\pi^-}(m^2_{\rho(1450)}-4m^2_\pi)^{3/2}} \nonumber\\
                         &=&  0.107.
   \label{th-Rho-1450}
\end{eqnarray}   
For the quasi-two-body decay $B^+\to \pi^+ [\rho(1450)^0\to] \pi^+\pi^-$, we have its branching fraction as 
$8.73^{+2.73}_{-2.54}\times 10^{-7}$ with the BW formula for $\rho(1450)^0$ and the relation 
$|c^K_{\rho(1450)}|  \approx  |c^\pi_{\rho(1450)}|$ in Eq.~(\ref{rela-ck1450}), where the error has the same sources as the branching 
fractions in Table~\ref{Res-1450} but have been added in quadrature.
This result are consistent with the measurements ${\mathcal B}=1.4^{+0.6}_{-0.9}\times10^{-6}$~\cite{prd79-072006,PDG-2020} 
from BaBar and ${\mathcal B}=(7.9\pm3.0)\times 10^{-7}$~\cite{prl124-031801,prd101-012006} by LHCb and agree with the 
prediction $(9.97\pm2.25)\times 10^{-7}$ in~\cite{prd101-111901} with the GS model for the resonance $\rho(1450)^0$. Then we 
have the ratio $R_{\rho(1450)}= 0.108^{+0.000}_{-0.001}$ which is very close to the $0.107$ in Eq.~(\ref{th-Rho-1450}) and 
the result in Fig.~\ref{fig-dep1450} for the ratio $R_{\rho(1450)}(\sqrt s)$ in the region around the mass of $\rho(1450)$ where the 
main portion of the branching fractions for $B^+\to \pi^+ [\rho(1450)^0\to] K^+K^-$ and $B^+\to \pi^+ [\rho(1450)^0\to] \pi^+\pi^-$ 
concentrated. The small error for $R_{\rho(1450)}$ from the PQCD predictions is caused by the cancellation, which means that the 
increase or decrease for the relevant numerical results from the uncertainties of those parameters will result in nearly identical change 
of the weight for these two decays. When the ${\rho(1450)^0}$ in Eq.~(\ref{th-Rho-1450}) is replaced by ${\rho(1700)^0}$, one will 
have the ratio $R_{\rho(1700)}\approx 0.143$~\cite{prd101-111901}. With the results ${\mathcal B}(\rho(1450)^0\to\pi^+\pi^-)=15\%$ 
and ${\mathcal B}(\rho(1700)^0\to\pi^+\pi^-)=14\%$ in Ref.~\cite{jhep2001-112} from CMD-3 Collaboration, one can estimate the 
branching fractions ${\mathcal B}(\rho(1450)^0\to K^+K^-)\approx1.6\%$ and ${\mathcal B}(\rho(1700)^0\to K^+K^-)\approx2.0\%$. 

It is important to notice that the definition of the coupling constant the Eq.~(\ref{def-cocont}) for the resonant states $\rho(770)$ and 
$\omega(782)$ decaying to the final state $K\bar K$ are invalid, or rather, one could not define the partial decay width such as  
$\Gamma_{\rho(770)\to K^+K^-}=\Gamma_{\rho(770)}{\mathcal B}_{\rho(770)^0\to K^+ K^-}$ or 
$\Gamma_{\omega(782)\to K^+K^-}=\Gamma_{\omega(782)}{\mathcal B}_{\omega(782)\to K^+ K^-}$ for the virtual contribution. 
This conclusion can be extended to other strong decay processes with the virtual contributions which come from the tails of the 
resonances.

In Ref.~\cite{prl123-231802}, the fit fraction of $\rho(1450)^0\to  K^+K^-$ for the three-body decays $B^\pm\to \pi^\pm K^+K^-$
was measured to be $(30.7\pm1.2\pm0.9)\%$ by LHCb Collaboration, implying ${\mathcal B}=(1.60\pm0.14)\times10^{-6}$ for the 
quasi-two-body decay $B^+\to\pi^+\rho(1450)^0 \to\pi^+ K^+K^-$~\cite{PDG-2020}. This branching fraction is close to the 
measurement ${\mathcal B}=1.4^{+0.6}_{-0.9}\times10^{-6}$ in~\cite{prd79-072006,PDG-2020} and larger than the result 
${\mathcal B}=(7.9\pm3.0)\times 10^{-7}$ from LHCb~\cite{prl124-031801,prd101-012006} for the 
$B^+\to \pi^+ [\rho(1450)^0\to] \pi^+\pi^-$ process. In view of the mass difference between kaon and pion, the factor 
$\left| \overrightarrow{q}\right|^3$ in Eq.~(\ref{eqn-diff-bra}) will be about $4.76$ times larger for the subprocess 
$\rho(1450)^0\to \pi^+\pi^-$ when comparing with $\rho(1450)^0\to K^+K^-$ for the decay $B^+\to\pi^+\rho(1450)^0$ at
$s=m^2_{\rho(1450)}$. It means that the coupling constant for $\rho(1450)^0\to K^+K^-$ should roughly be $\sqrt{4.76}$ times
larger than that for $\rho(1450)^0\to \pi^+\pi^-$ in order to achieve the comparable branching fractions for the quasi-two-body 
decays $B^+\to\pi^+[\rho(1450)^0 \to] K^+K^-$ and $B^+\to \pi^+ [\rho(1450)^0\to] \pi^+\pi^-$. Clearly, a larger coupling constant 
for $\rho(1450)^0\to K^+K^-$ is contrary to the naive expectation~\cite{2007-02558} and the discussions in literature~\cite{epjc39-41,
plb779-64}.

\section{Summary}  \label{sec-con}  

In this work, we studied the contributions for kaon pair originating from the resonances $\rho(770)$, $\omega(782)$ and their 
excited states $\rho(1450,1700)$ and $\omega(1420,1650)$ in the three-body decays $B\to K\bar K h$ in the PQCD approach.
The subprocesses $\rho(770,1450,1700)\to K\bar K$ and $\omega(782,1420,1650)\to K\bar K$, which can not be calculated in 
the PQCD, were introduced into the distribution amplitudes for $K\bar K$ system via the kaon vector time-like form factors. 
With the coefficients $c^K_{\rho(770)}=1.247\pm0.019$, $c^K_{\omega(782)}=1.113\pm0.019$, $c^K_{\rho(1450)}=-0.156\pm0.015$, 
$c^K_{\omega(1420)}=-0.117\pm0.013$ and $c^K_{\omega(1650),\rho(1700)}=-0.083\pm0.019$ in the time-like form factors 
for kaons, we predicted the $CP$ averaged branching fractions and the direct $CP$ asymmetries for the quasi-two-body 
processes $B\to \rho(770,1450,1700) h\to K\bar K h$ and $B\to\omega(782,1420,1650) h \to K\bar K h$.

The branching fractions of the virtual contributions for $K\bar K$ in this work from the BW tails of the intermediate states 
$\rho(770)$ and $\omega(782)$ in the concerned decays which have been ignored in experimental and theoretical studies were 
found larger than the corresponding results from $\rho(1450,1700)$ and $\omega(1420,1650)$. A remarkable phenomenon 
for the virtual contributions discussed in this work is that the differential branching fractions for $B\to \rho(770) h\to K\bar K h$ 
and $B\to\omega(782) h \to K\bar K h$ are nearly unaffected by the quite different values of the full widths for $\rho(770)$ and 
$\omega(782)$. The definition of the partial decay width such as 
$\Gamma_{\rho(770)\to K^+K^-}=\Gamma_{\rho(770)}{\mathcal B}_{\rho(770)^0\to K^+ K^-}$ for the virtual contribution are 
invalid. This conclusion can be extended to other strong decay processes with the virtual contributions come from the tails 
of the resonances. The bumps of the lines for the differential branching fractions for those virtual contributions, which are 
generated by the phase space factors and the tails of the BW formula of $\rho(770)$ or $\omega(782)$, should not be taken 
as the evidence for a new resonant state at about $1.35$ GeV.

The PQCD predicted results for the branching fractions of the quasi-two-body decays $B^+\to \pi^+ \rho(1450)^0\to \pi^+K^+K^-$ 
and $B^+\to \pi^+ \rho(1450)^0\to \pi^+\pi^+\pi^-$ meet the requirement of the $SU(3)$ symmetry relation 
$g_{\rho(1450)^0K^+K^-}\approx \frac12 g_{\rho(1450)^0\pi^+\pi^-}$. The larger coupling constant for $\rho(1450)^0\to K^+K^-$ 
deduced from the fit fraction $(30.7\pm1.2\pm0.9)\%$ for $\rho(1450)^0\to  K^+K^-$ in the $B^\pm\to \pi^\pm K^+K^-$ decays 
by LHCb Collaboration is contrary to the discussions in literature. We estimated the branching fractions to be about $1.6\%$ and 
$2.0\%$ for the decays $\rho(1450)^0\to K^+K^-$ and $\rho(1700)^0\to K^+K^-$, respectively, according to the measurement 
results from CMD-3 Collaboration for $\rho(1450,1700)^0\to\pi^+\pi^-$.

\begin{acknowledgments}
This work was supported in part by the National Natural Science Foundation of China 
under Grants No.~11547038 and No.~11575110.
\end{acknowledgments}                 

\appendix
\section{Distribution amplitudes} \label{sec-DAs}

The $B$ meson light-cone matrix element can be decomposed as~\cite{npb592-3,prd76-074018}
\begin{eqnarray}
\Phi_B=\frac{i}{\sqrt{2N_c}} (p{\hspace{-1.8truemm}/}_B+m_B)\gamma_5\phi_B (k_B),
\label{bmeson}
\end{eqnarray}
where the distribution amplitude $\phi_B$ is of the form
\begin{eqnarray}
\phi_B(x_B,b_B)= N_B x_B^2(1-x_B)^2
\mathrm{exp}\left[-\frac{(x_Bm_B)^2}{2\omega_{B}^2} -\frac{1}{2} (\omega_{B}b_B)^2\right],
\label{phib}
\end{eqnarray}
with $N_B$ the normalization factor. The shape parameters $\omega_B = 0.40 \pm 0.04$ GeV for $B^\pm$ and $B^0$ and 
$\omega_{B_s}=0.50 \pm 0.05$ for $B^0_s$, respectively.

The light-cone wave functions for pion and kaon are written as~\cite{jhep9809-005,jhep9901-010,prd71-014015,jhep0605-004}
\begin{eqnarray}
\Phi_{h}=\frac{i}{\sqrt{2N_c}}\gamma_5\left[p{\hspace{-1.8truemm}/}_3\phi^A(x_3)+m^h_0\phi^P(x_3)+
m^h_0(n\hspace{-2.0truemm}/ v\hspace{-1.8truemm}/-1)\phi^T(x_3)\right].
\end{eqnarray}
The distribution amplitudes of $\phi^A(x_3), \phi^P(x_3)$ and $\phi^T(x_3)$ are
\begin{eqnarray}
\phi^A(x_3)&=&\frac{f_{h}}{2\sqrt{2N_c}}6x_3(1-x_3)\left[1+a_1^{h}C_1^{3/2}(t)+a_2^{h}C_2^{3/2}(t)+a_4^{h}C_4^{3/2}(t)\right], \\
\phi^P(x_3)&=&\frac{f_{h}}{2\sqrt{2N_c}}\left[1+(30\eta_3-\frac{5}{2}\rho^2_{h})C_2^{1/2}(t)
-3\big[\eta_3\omega_3+\frac{9}{20}\rho^2_{h}(1+6a_2^{h})\big]C_4^{1/2}(t)\right], \\
\phi^T(x_3)&=&\frac{f_{h}}{2\sqrt{2N_c}}(-t)\left[1+6\left(5\eta_3-\frac{1}{2}\eta_3\omega_3-\frac{7}{20}\rho^2_{h}
-\frac{3}{5}\rho^2_{h}a_2^{h}\right)(1-10x_3+10x_3^2)\right],\quad\;
\end{eqnarray}
with $t=2x_3-1$, $C^{1/2}_{2,4}(t)$ and $C^{3/2}_{1,2,4}(t)$ are Gegenbauer polynomials. 
The chiral scale parameters $m^h_0=\frac{m^2_{h}}{m_q+m_{q^\prime}}$ for pion and kaon are 
$m_0^{\pi}=(1.4 \pm 0.1)$ GeV and $m_0^{K}=(1.9 \pm 0.1)$ GeV as they in~\cite{prd76-074018}.
The decay constants $f_\pi=130.2(1.2)$ MeV and $f_K=155.7(3)$ MeV can be found in Ref.~\cite{PDG-2020}. 
The Gegenbauer moments $a_1^{\pi}=0, a_1^K=0.06, a_2^{h}=0.25, a_4^{h}=-0.015$ and the parameters 
$\rho_{h}=m_{h}/m_0^{h}, \eta_3=0.015, \omega_3=-3$ are adopted in the numerical calculation.

\section{Decay amplitudes} \label{sec-decayamp}

With the subprocesses $\rho^+\to K^+ \bar K^0 $, $\rho^-\to K^- K^0$, $\rho^0 \to K^+ K^-$, $\rho^0 \to K^0\bar K^0$, 
$\omega \to K^+ K^-$ and $\omega \to K^0\bar K^0$, and $\rho$ is $\rho(770), \rho(1450)$ or $\rho(1700)$ and $\omega$ 
is $\omega(782), \omega(1420)$ or $\omega(1650)$, the Lorentz invariant decay amplitudes for the quasi-two-body decays 
$B\to \rho h\to K\bar K h$ and $B\to\omega h \to K\bar K h$ are given as follows:
\begin{eqnarray}
 {\cal A}(B^+ \to \rho^+\pi^0) 
   &=&\frac{G_F}{2} V_{ub}^*V_{ud}\big\{a_1[F^{LL}_{Th}+F^{LL}_{Ah}-F^{LL}_{a\rho}]+a_2F^{LL}_{T\rho}
               +C_1[M^{LL}_{Th}+M^{LL}_{Ah}      \nonumber\\
   &&-M^{LL}_{A\rho}]   +C_2 M^{LL}_{T\rho}\big\}- \frac{G_F}{2} V_{tb}^*V_{td} \big\{ \big[-a_4+\frac{5 C_9}{3}+C_{10}
              -\frac{3 a_7}{2}\big] F^{LL}_{T\rho}          \nonumber\\
   &&- [a_6-\frac{a_8}{2}] F^{SP}_{T\rho}+ [\frac{C_9+3 C_{10}}{2}-C_3] M^{LL}_{T\rho}-[C_5-\frac{C_7}{2}] M^{LR}_{T\rho}  
              \nonumber\\
   &&+\frac{3 C_8}{2}M^{SP}_{T\rho} +[a_4+a_{10}][F^{LL}_{Th}+F^{LL}_{Ah}-F^{LL}_{A\rho}]
              +[a_6+a_8][F^{SP}_{Ah}                     \nonumber\\
   &&- F^{SP}_{A\rho}]+[C_3+C_9][M^{LL}_{Th}+M^{LL}_{Ah}-M^{LL}_{A\rho}]+[C_5+C_7][M^{LR}_{Th}                   
              \nonumber\\
   &&+ M^{LR}_{Ah}-M^{LR}_{A\rho}]\big\},       \label{amp1}
\end{eqnarray}
\begin{eqnarray}
 {\cal A}(B^+ \to \rho^0\pi^+) 
   &=& \frac{G_F}{2} V_{ub}^*V_{ud} \big\{ a_1[F^{LL}_{T\rho}+F^{LL}_{A\rho}-F^{LL}_{Ah}]
            +a_2F^{LL}_{Th}+C_1[M^{LL}_{T\rho}+M^{LL}_{A\rho}    \nonumber\\
  &&-M^{LL}_{Ah}] +C_2 M^{LL}_{Th}\big\}- \frac{G_F}{2} V_{tb}^*V_{td}\big\{ [a_4+a_{10}][F^{LL}_{T\rho}+F^{LL}_{A\rho}
           -F^{LL}_{Ah}]    \nonumber\\
  &&+[a_6+a_8][F^{SP}_{T\rho}+F^{SP}_{A\rho}-F^{SP}_{Ah}]+[C_3+C_9][M^{LL}_{T\rho}+M^{LL}_{A\rho}-M^{LL}_{Ah}]  
       \nonumber\\
  &&+[C_5+C_7][M^{LR}_{T\rho}+M^{LR}_{A\rho}-M^{LR}_{Th}]+[\frac{5}{3}C_9+C_{10}+\frac{3 a_7}{2}-a_4]F^{LL}_{Th}  
       \nonumber\\
  &&+[\frac{C_9+3 C_{10}}{2}-C_3]M^{LL}_{Th}-[C_5-\frac{C_7}{2}]M^{LR}_{Th}+\frac{3 C_8}{2}M^{SP}_{Th}\big \},      
      \label{amp2}
\end{eqnarray}
\begin{eqnarray}
 {\cal A}(B^+ \to \omega\pi^+) 
   &=& \frac{G_F}{2} V_{ub}^*V_{ud}\big\{ a_1[F^{LL}_{T\omega}+F^{LL}_{A\omega}+F^{LL}_{Ah}]+a_2F^{LL}_{Th}
             +C_1[M^{LL}_{T\omega}+M^{LL}_{A\omega}   \nonumber\\
  &&+M^{LL}_{Ah}] +C_2 M^{LL}_{Th}\big\}-\frac{G_F}{2}V_{tb}^*V_{td}\big\{ [a_4+a_{10}][F^{LL}_{T\omega}+F^{LL}_{A\omega}
            +F^{LL}_{Ah}]       \nonumber\\
  &&+[a_6+a_8][F^{SP}_{T\omega}+F^{SP}_{A\omega}+F^{SP}_{Ah}]+[C_3+C_9][M^{LL}_{T\omega}+M^{LL}_{A\omega}
          +M^{LL}_{Ah}]        \nonumber\\
  &&+[C_5+C_7][M^{LR}_{T\omega}+M^{LR}_{A\omega}+M^{LR}_{Ah}]+[(7 C_3+5 C_4+C_9-C_{10})/3           \nonumber\\
  &&+2 a_5+\frac{a_7}{2}]F^{LL}_{Th}+[C_3+2 C_4-\frac{C_9-C_{10}}{2}]M^{LL}_{Th}+[C_5-\frac{C_7}{2}]M^{LR}_{Th}   \nonumber\\
  &&+[2 C_6+\frac{C_8}{2}]M^{SP}_{Th}\big\},      \label{amp3} 
\end{eqnarray}
\begin{eqnarray}
{\cal A}(B^+ \to \rho^+K^0) 
 &=& \frac{G_F}{\sqrt{2}} V_{ub}^*V_{us}\{a_1F^{LL}_{A\rho}+C_1 M^{LL}_{A\rho}\}
         - \frac{G_F}{\sqrt{2}}V_{tb}^*V_{ts}\big\{[a_4-\frac{a_{10}}{2}]F^{LL}_{T\rho} +[a_6     \nonumber\\
 &&-\frac{a_8}{2}] F^{SP}_{T\rho}+[C_3-\frac{C_9}{2}]M^{LL}_{T\rho}+[C_5-\frac{C_7}{2}]M^{LR}_{T\rho}
       +[a_4+a_{10}]F^{LL}_{A\rho}              \nonumber\\
 &&+[C_3+C_9]M^{LL}_{A\rho}+[a_6+a_8]F^{SP}_{A\rho}+[C_5+C_7]M^{LR}_{A\rho}\big\},
          \label{amp4}
\end{eqnarray}
\begin{eqnarray}
{\cal A}(B^+ \to \rho^0K^+) 
  &=& \frac{G_F}{2} V_{ub}^*V_{us}\big\{ a_1[F^{LL}_{T\rho}+F^{LL}_{A\rho}]+a_2 F^{LL}_{Th}+C_1[M^{LL}_{T\rho}
          +M^{LL}_{A\rho}]+C_2M^{LL}_{Th}\big\}       \nonumber\\
  && -\frac{G_F}{2} V_{tb}^*V_{ts}\big\{[a_4+a_{10}][F^{LL}_{T\rho}+F^{LL}_{A\rho}]+[a_6+a_8][F^{SP}_{T\rho}
         +F^{SP}_{A\rho}]+[C_3                                  \nonumber\\
  && +C_9][M^{LL}_{T\rho}+M^{LL}_{A\rho}]+[C_5+C_7][M^{LR}_{T\rho}+M^{LR}_{A\rho}]  +\frac{3}{2}[a_7+a_9]F^{LL}_{Th} 
                                                                                \nonumber\\
  &&+\frac{3 C_{10} }{2} M^{LL}_{Th} +\frac{3 C_8}{2} M^{SP}_{Th}\big\}, \label{amp5}
\end{eqnarray}
\begin{eqnarray}
{\cal A}(B^+ \to \omega K^+) 
  &=& \frac{G_F}{2} V_{ub}^*V_{us}\big\{ a_1[F^{LL}_{T\omega}+F^{LL}_{A\omega}]+a_2 F^{LL}_{Th}
               +C_1[M^{LL}_{T\omega}+M^{LL}_{A\omega}] +C_2M^{LL}_{Th}\big\}          \nonumber\\
  && - \frac{G_F}{2} V_{tb}^*V_{ts}\big\{[a_4+a_{10}][F^{LL}_{T\omega}+F^{LL}_{A\omega}]
             +[a_6+a_8][F^{SP}_{T\omega}+F^{SP}_{A\omega}] +[C_3                              \nonumber\\
  &&+C_9][M^{LL}_{T\omega}+M^{LL}_{A\omega}]+[C_5+C_7][M^{LR}_{T\omega}+M^{LR}_{A\omega}]+[2 a_3+2 a_5+a_7/2 
                   			\nonumber\\
  &&+a_9/2]F^{LL}_{Th}+[2 C_4+\frac{C_{10}}{2}] M^{LL}_{Th} +[2 C_6+\frac{C_8}{2}] M^{SP}_{Th}\big\}, 
        \label{amp6}
\end{eqnarray}
\begin{eqnarray}
{\cal A}(B^0 \to \rho^+\pi^-) 
  &=& \frac{G_F} {\sqrt{2}} V_{ub}^*V_{ud}\big\{a_2F^{LL}_{A\rho}+C_2 M^{LL}_{A\rho}+a_1F^{LL}_{Th}+C_1 M^{LL}_{Th}\big\}
             -\frac{G_F} {\sqrt{2}} V_{tb}^*V_{td}\big\{[a_3         \nonumber\\
  &&+a_9-a_5-a_7]F^{LL}_{A\rho}+[C_4+C_{10}]M^{LL}_{A\rho}+[C_6+C_8]M^{SP}_{A\rho}+[a_4+a_{10}]  \nonumber\\
  &&\times F^{LL}_{Th} +[C_3+C_9]M^{LL}_{Th}+[C_5+C_7]M^{LR}_{Th}+[\frac{4}{3}[C_3+C_4-\frac{C_9}{2}-\frac{C_{10}}{2}]
   											\nonumber\\
  &&-a_5+\frac{a_7}{2}]F^{LL}_{Ah} +[a_6-\frac{a_8}{2}]F^{SP}_{Ah}+[C_3+C_4-\frac{C_9}{2}
            -\frac{C_{10}}{2}]M^{LL}_{Ah} +[C_5					\nonumber\\
  &&-\frac{C_7}{2}]M^{LR}_{Ah}+[C_6-\frac{C_8}{2}]M^{SP}_{Ah}\big\},
   	\label{amp7}
\end{eqnarray}
\begin{eqnarray}
{\cal A}(B^0 \to \rho^-\pi^+) 
  &=& \frac{G_F} {\sqrt{2}}V_{ub}^*V_{ud}\big\{ a_1 F^{LL}_{T\rho}+a_2 F^{LL}_{Ah}+C_1 M^{LL}_{T\rho}+C_2 M^{LL}_{Ah}\big\}
   -\frac{G_F} {\sqrt{2}} V_{tb}^*V_{td}\big\{[a_4		\nonumber\\
  &&+a_{10}]F^{LL}_{T\rho}+[a_6+a_8]F^{SP}_{T\rho}+[C_3+C_9]M^{LL}_{T\rho}+[C_5+C_7]M^{LR}_{T\rho}	\nonumber\\
  &&+[\frac{4}{3}[C_3+C_4-\frac{C_9+C_{10}}{2}]-a_5+\frac{a_7}{2}]F^{LL}_{A\rho}+[a_6-\frac{a_8}{2}]F^{SP}_{A\rho}	\nonumber\\
  &&+[C_3+C_4-\frac{C_9+C_{10}}{2}]M^{LL}_{A\rho}+[C_5-\frac{C_7}{2}]M^{LR}_{A\rho}+[C_6-\frac{C_8}{2}]M^{SP}_{A\rho}
  				\nonumber\\
  &&+[a_3+a_9-a_5-a_7]F^{LL}_{Ah}+[C_4+C_{10}]M^{LL}_{Ah}+[C_6+C_8]M^{SP}_{aP}\big\},
        \label{amp8}
\end{eqnarray}
\begin{eqnarray}
{\cal A}(B^0 \to \rho^0\pi^0) 
  &=& \frac{G_F} {2\sqrt{2}} V_{ub}^*V_{ud}\big\{a_2[F^{LL}_{A\rho}+F^{LL}_{Ah}-F^{LL}_{T\rho}-F^{LL}_{Th}]
        +C_2[M^{LL}_{A\rho}+M^{LL}_{Ah}		\nonumber\\
  &&-M^{LL}_{T\rho}-M^{LL}_{Th}]\big\}-\frac{G_F} {2\sqrt{2}} V_{tb}^*V_{td}\big\{[a_4-\frac{5 C_9}{3}-C_{10}
        +\frac{3 a_7}{2}]F^{LL}_{T\rho} +[a_6		\nonumber\\
  &&-\frac{a_8}{2}][F^{SP}_{T\rho}	+F^{SP}_{A\rho}+F^{SP}_{Ah}]+[C_3-\frac{C_9+3 C_{10}}{2}][M^{LL}_{T\rho}+M^{LL}_{Th}]+[C_5
                  \nonumber\\
  &&-\frac{C_7}{2}][M^{LR}_{T\rho}+M^{LR}_{A\rho}+M^{LR}_{Th}+M^{LR}_{Ah}]-\frac{3 C_8}{2}[M^{SP}_{T\rho}+M^{SP}_{Th}] 
        +[(7 C_3      \nonumber\\
  &&+5 C_4+C_9 -C_{10})/{3}-2 a_5-\frac{a_7}{2}][F^{LL}_{A\rho}+F^{LL}_{Ah}]+[C_3+2 C_4
                 \nonumber\\
  &&-\frac{C_9- C_{10}}{2}] [M^{LL}_{A\rho}+M^{LL}_{Ah}]+[2 C_6+\frac{C_8}{2}][M^{SP}_{A\rho}+M^{SP}_{Ah}]+[a_4-\frac{5 C_9}{3}
               \nonumber\\
  &&-C_{10}-\frac{3 a_7}{2}] F^{LL}_{Th}\big\},
               \label{amp9}
\end{eqnarray}
\begin{eqnarray}
{\cal A}(B^0 \to \omega\pi^0) 
  &=& \frac{G_F} {2\sqrt{2}} V_{ub}^*V_{ud}\big\{a_2[F^{LL}_{A\omega}+F^{LL}_{Ah}+F^{LL}_{T\omega}-F^{LL}_{Th}]
       +C_2[M^{LL}_{A\omega}+M^{LL}_{Ah}  +M^{LL}_{T\omega}		\nonumber\\
  &&-M^{LL}_{Th}]\big\}-\frac{G_F} {2\sqrt{2}}V_{tb}^*V_{td}\big\{[-a_4+\frac{5 C_9}{3}+C_{10}
       -\frac{3 a_7}{2}][F^{LL}_{T\omega}+F^{LL}_{A\omega}+F^{LL}_{Ah}]		\nonumber\\
  &&-[a_6-\frac{a_8}{2}][F^{SP}_{T\omega}+F^{SP}_{A\omega}+F^{SP}_{Ah}]-[(7 C_3+5 C_4+C_9-C_{10})/{3}+2 a_5   \nonumber\\
  &&+\frac{a_7}{2}] F^{LL}_{Th}-[C_3-\frac{C_9+3 C_{10}}{2}][M^{LL}_{T\omega}+M^{LL}_{A\omega}+M^{LL}_{Ah}]
       -[C_5-\frac{C_7}{2}][M^{LR}_{T\omega}				\nonumber\\
  &&+M^{LR}_{A\omega}+M^{LR}_{Th}+M^{LR}_{Ah}]+\frac{3 C_8}{2}[M^{SP}_{T\omega}+M^{SP}_{A\omega}+M^{SP}_{Ah}]
     -[C_3+2 C_4			\nonumber\\
  &&-\frac{C_9- C_{10}}{2}]M^{LL}_{Th}-[2 C_6+\frac{C_8}{2}]M^{SP}_{Th}\big\},		\label{amp10}
\end{eqnarray}
\begin{eqnarray}
{\cal A}(B^0 \to \rho^-K^+) 
  &=& \frac{G_F} {\sqrt{2}}V_{ub}^*V_{us}\big\{a_1 F^{LL}_{T\rho}+C_1 M^{LL}_{T\rho}\big\} 
         -\frac{G_F} {\sqrt{2}}V_{tb}^*V_{ts}\big\{[a_4+a_{10}]F^{LL}_{T\rho}+[a_6+a_8]  		\nonumber\\
  &&\times F^{SP}_{T\rho} +[C_3+C_9]M^{LL}_{T\rho}+[C_5+C_7]M^{LR}_{T\rho}+[a_4-\frac{a_{10}}{2}]F^{LL}_{A\rho} 
       +[a_6-\frac{a_8}{2}]  			\nonumber\\
  &&\times F^{SP}_{A\rho}+[C_3-\frac{C_9}{2}]M^{LL}_{A\rho} +[C_5-\frac{C_7}{2}]M^{LR}_{A\rho}\big\},
       		\label{amp11}
\end{eqnarray}
\begin{eqnarray}
{\cal A}(B^0 \to \rho^0K^0) 
  &=& \frac{G_F} {2}V_{ub}^*V_{us}\big\{a_2 F^{LL}_{Th}+C_2 M^{LL}_{Th}]\big\}\!
        -\frac{G_F} {2}V_{tb}^*V_{ts}\big\{\!-\![a_4-\frac{a_{10}}{2}][F^{LL}_{T\rho}+F^{LL}_{A\rho}]		\nonumber\\
  &&-[a_6-\frac{a_8}{2}][F^{SP}_{T\rho}+F^{SP}_{A\rho}]-[C_3-\frac{C_9}{2}][M^{LL}_{T\rho}+M^{LL}_{A\rho}]
       -[C_5-\frac{C_7}{2}]		\nonumber\\
  &&\times [M^{LR}_{T\rho}+M^{LR}_{A\rho}]+\frac{3}{2}[a_7+a_9]F^{LL}_{Th}+\frac{3 C_{10}}{2}M^{LL}_{Th}
      +\frac{3 C_8}{2}M^{SP}_{Th}\big\},
                 \label{amp12}
\end{eqnarray}
\begin{eqnarray}{\cal A}(B^0 \to \omega K^0) 
  &=& \frac{G_F} {2}V_{ub}^*V_{us}\big\{a_2 F^{LL}_{Th}+C_2 M^{LL}_{Th}\big\}
          - \frac{G_F} {2}V_{tb}^*V_{ts}\big\{[a_4-\frac{a_{10}}{2}][F^{LL}_{T\omega}+F^{LL}_{A\omega}]    	  \nonumber\\
  &&+[a_6-\frac{a_8}{2}][F^{SP}_{T\omega}+F^{SP}_{A\omega}]+[C_3-\frac{C_9}{2}][M^{LL}_{T\omega}+M^{LL}_{A\omega}]
         +[C_5-\frac{C_7}{2}] 	  \nonumber\\
  &&\times [M^{LR}_{T\omega}+M^{LR}_{A\omega}]+[2 a_3+2 a_5+\frac{a_7+a_9}{2}]F^{LL}_{Th}+[2 C_4+\frac{C_{10}}{2}]M^{LL}_{Th} 
		          \nonumber\\
  &&+[2 C_6+\frac{C_8}{2}]M^{SP}_{Th}\big\},		\label{amp13}
\end{eqnarray}
\begin{eqnarray}
{\cal A}(B_s^0 \to \rho^+\pi^-) 
  &=& \frac{G_F} {\sqrt{2}} V_{ub}^*V_{us}\big\{a_2 F^{LL}_{A\rho}+C_2 M^{LL}_{A\rho}\big\}
        -\frac{G_F} {\sqrt{2}}V_{tb}^*V_{ts}\big\{[a_3+a_9-a_5-a_7]F^{LL}_{A\rho}			\nonumber\\
  &&+[C_4+C_{10}]M^{LL}_{A\rho}+[C_6+C_8]M^{SP}_{A\rho}+[a_3-\frac{a_9}{2}-a_5+\frac{a_7}{2}]F^{LL}_{Ah}+[C_4      \nonumber\\
  &&-\frac{C_{10}}{2}]M^{LL}_{Ah}+[C_6-\frac{C_8}{2}]M^{SP}_{Ah}\big\},		\label{amp14}
\end{eqnarray}
\begin{eqnarray}
{\cal A}(B_s^0 \to \rho^-\pi^+) 
  &=& \frac{G_F} {\sqrt{2}}V_{ub}^*V_{us}\big\{a_2 F^{LL}_{Ah}+C_2 M^{LL}_{Ah}\big\}
        -\frac{G_F} {\sqrt{2}}V_{tb}^*V_{ts}\big\{[a_3-\frac{a_9}{2}
          -a_5+\frac{a_7}{2}]F^{LL}_{A\rho}		\nonumber\\
  &&+[C_4-\frac{C_{10}}{2}]M^{LL}_{A\rho}+[C_6-\frac{C_8}{2}]M^{SP}_{A\rho}+[a_3+a_9-a_5-a_7]F^{LL}_{Ah}+[C_4 	\nonumber\\
  &&+C_{10}]M^{LL}_{Ah}+[C_6+C_8]M^{SP}_{Ah}\big\},\label{amp15}
\end{eqnarray}
\begin{eqnarray}
{\cal A}(B_s^0 \to \rho^0\pi^0) 
  &=& \frac{G_F} {2\sqrt{2}} V_{ub}^*V_{us}\big\{a_2[F^{LL}_{A\rho}+F^{LL}_{Ah}]+C_2[M^{LL}_{A\rho}+M^{LL}_{Ah}]\big\}
          -\frac{G_F} {2\sqrt{2}} V_{tb}^*V_{ts}\big\{[2 a_3       \nonumber\\
  &&+\frac{a_9}{2}-2 a_5-\frac{a_7}{2}][F^{LL}_{A\rho}+F^{LL}_{Ah}]+[2C_4+\frac{C_{10}}{2}][M^{LL}_{A\rho}
        +M^{LL}_{Ah}]+[2C_6			\nonumber\\
  &&+\frac{C_8}{2}][M^{SP}_{A\rho}+M^{SP}_{Ah}]\big\},
  \label{amp16}
\end{eqnarray}
\begin{eqnarray}
{\cal A}(B_s^0 \to \omega \pi^0) 
   &=& \frac{G_F} {2\sqrt{2}}V_{ub}^*V_{us}\big\{a_2[F^{LL}_{A\omega}+F^{LL}_{Ah}]+C_2[M^{LL}_{A\omega}
         +M^{LL}_{Ah}]\big\}-\frac{G_F} {2\sqrt{2}}V_{tb}^*V_{ts}			\nonumber\\
   &&\times\big\{\frac{3}{2}[a_9-a_7][F^{LL}_{A\omega}+F^{LL}_{Ah}]+\frac{3 C_{10}}{2}[M^{LL}_{A\omega}+M^{LL}_{Ah}]
         +\frac{3 C_8}{2}[M^{SP}_{A\omega}			\nonumber\\
   &&+M^{SP}_{Ah}]\big\},
   \label{amp17}
\end{eqnarray}
\begin{eqnarray}
{\cal A}(B_s^0 \to \rho^+K^-) 
  &=&  \frac{G_F} {\sqrt{2}} V_{ub}^*V_{ud}\big\{a_1 F^{LL}_{Th}+C_1 M^{LL}_{Th}\big\}
        -\frac{G_F} {\sqrt{2}}V_{tb}^*V_{td}\big\{[a_4+a_{10}]F^{LL}_{Th}+[C_3		\nonumber\\
  &&+C_9] M^{LL}_{Th}+[C_5+C_7]M^{LR}_{Th}+[a_4-\frac{a_{10}}{2}]F^{LL}_{Ah}+[a_6-\frac{a_8}{2}]F^{SP}_{Ah}		\nonumber\\
  &&+[C_3-\frac{C_9}{2}]M^{LL}_{Ah}+[C_5-\frac{C_7}{2}]M^{LR}_{Ah}\big\},
     \label{amp18}
\end{eqnarray}  
\begin{eqnarray}
{\cal A}(B_s^0 \to \rho^0 \bar K^0) 
  &=& \frac{G_F} {2} V_{ub}^*V_{ud}\big\{a_2 F^{LL}_{Th}+C_2 M^{LL}_{Th}\big\}
           -\frac{G_F} {2} V_{tb}^*V_{td}\big\{ [\frac{5 C_9}{3}+C_{10}+\frac{3a_7}{2}-a_4]         \nonumber\\
  &&\times F^{LL}_{Th} +[\frac{C_9}{2}+\frac{3 C_{10}}{2}-C_3]M^{LL}_{Th}-[C_5-\frac{C_7}{2}][M^{LR}_{Th}+M^{LR}_{Ah}]
          +\frac{3 C_8}{2}  	\nonumber\\
  &&\times M^{SP}_{Th}-[a_4-\frac{a_{10}}{2}]F^{LL}_{Ah}-[a_6-\frac{a_8}{2}]F^{SP}_{Ah}-[C_3-\frac{C_9}{2}]M^{LL}_{Ah}\big\},
        \label{amp19}
\end{eqnarray}
\begin{eqnarray}
{\cal A}(B_s^0 \to \omega \bar K^0) 
  &=& \frac{G_F} {2}V_{ub}^*V_{ud}\big\{a_2 F^{LL}_{Th}+C_2 M^{LL}_{Th}\big\}
           -\frac{G_F} {2} V_{tb}^*V_{td}\big\{ [(7 C_3+5 C_4+C_9		\nonumber\\
  &&-C_{10})/{3} +2 a_5+\frac{a_7}{2}]F^{LL}_{Th}+[C_3+2 C_4-\frac{C_9-C_{10}}{2}]M^{LL}_{Th} +[C_5-\frac{C_7}{2}]    \nonumber\\
  &&\times [M^{LR}_{Th}+M^{LR}_{Ah}]+[2 C_6+\frac{C_8}{2}]M^{SP}_{Th}+[a_4-\frac{a_{10}}{2}]F^{LL}_{Ah} 
          +[a_6-\frac{a_8}{2}]F^{SP}_{Ah}   \nonumber\\
  &&+[C_3-\frac{C_9}{2}]M^{LL}_{Ah}\big\},	\label{amp20}
\end{eqnarray}  
where $G_F$ is the Fermi coupling constant, $V$'s are the CKM matrix elements. The combinations $a_i$ with $i=1$-$10$ 
are defined as
\begin{eqnarray}
  &~& a_1=C_2+{C_1}/{3}, ~~a_2= C_1+{C_2}/{3}, ~~a_3= C_3+{C_4}/{3}, ~~a_4=C_4+{C_3}/{3},    \nonumber\\
  &~& a_5= C_5+ {C_6}/{3}, ~~a_6= C_6+{C_5}/{3}, ~~a_7=C_7+{C_8}/{3}, ~~a_8= C_8+{C_7}/{3},  \nonumber\\
  &~& a_9= C_9+{C_{10}}/{3}, ~a_{10}= C_{10}+{C_{9}}/{3},
\end{eqnarray}
for the Wilson coefficients. 

The general amplitudes for the quasi-two body decays $B\to \rho h \to K\bar K h$ and $B\to \omega h \to K\bar K h$ in the 
decay amplitudes Eqs.~(\ref{amp1})-(\ref{amp20}) are given according to Fig.~\ref{fig-feyndiag}, the typical Feynman 
diagrams for the PQCD approach. The symbols $LL$, $LR$ and $SP$ are employed to denote the amplitudes from the 
$(V-A)(V-A)$, $(V-A)(V+A)$ and $(S-P)(S+P)$ operators, respectively. The emission diagrams are depicted in 
Fig.~\ref{fig-feyndiag} (a) and (c), while the annihilation diagrams are shown by Fig.~\ref{fig-feyndiag} (b) and (d).  
For the factorizable diagrams in Fig.~\ref{fig-feyndiag}, we name their expressions with $F$, while the others are nonfactorizable 
diagrams, we name their expressions with $M$. The specific  expressions for these general amplitudes 
are the same as in the appendix of~\cite{epjc80-815} but with the replacements $\phi\to\rho$ and $\phi\to\omega$ for their subscripts 
for the subprocesses $\rho\to K\bar K$ and $\omega\to K\bar K$, respectively, in this work.  It should be understood that the Wilson 
coefficients $C$ and the amplitudes $F$ and $M$ for the factorizable and nonfactorizable contributions, respectively, appear in 
convolutions in momentum fractions and impact parameters $b$.


\end{document}